\documentclass[runningheads]{llncs}
\usepackage[T1]{fontenc}
\usepackage{makeidx}  

\usepackage{amssymb,amsmath,amsfonts}
\usepackage{pifont}
\usepackage{graphicx}
\usepackage{latexsym}
\usepackage{todonotes}
\usepackage{xspace}
\usepackage{tikz,pgf}
\usepackage{paralist}
\usepackage{xcolor}
\usepackage{wasysym}
\usepackage{wrapfig}
\usepackage[clock]{ifsym}
\usepackage{orcidlink}

\usetikzlibrary{automata,calc,arrows,shapes,backgrounds}

\definecolor{ly}{RGB}{250,250,200}
\definecolor{lg}{RGB}{200,250,200}

\def\Nset{\mathbb{N}}

\def\Qset{\mathbb{Q}}
\def\Rset{\mathbb{R}}

\def\ra{\rightarrow}

\newcommand{\sat}{\models}

\newcommand{\propref}[1]{Proposition~\ref{#1}}

\newcommand{\lemref}[1]{Lemma~\ref{#1}}
\newcommand{\figref}[1]{Figure~\ref{#1}}

\newcommand{\setsep}{\, \mid \,}

\newcommand{\IGNORE}[1]{}

\newcommand{\nnr}{\Rset_{\geq 0}}

\newcommand{\dist}{\mathsf{Dist}}
\newcommand{\adist}{\mu}

\newcommand{\support}{\mathsf{support}}

\newcommand{\anMDP}{\mathcal{M}}
\newcommand{\mdptrans}{\Delta}

\newcommand{\astate}{s}

\newcommand{\sinit}{\overline{\astate}}

\newcommand{\anaction}{a}

\newcommand{\afinrun}{r}

\newcommand{\last}{\mathit{last}}

\newcommand{\astrat}{\sigma}

\newcommand{\clocks}{\mathcal{X}}
\newcommand{\aclock}{x}
\newcommand{\anotherclock}{y}

\newcommand{\aval}{v}
\newcommand{\valuations}{\nnr^{\clocks}}
\newcommand{\aclockset}{X}

\newcommand{\pta}{\mathcal{P}}
\newcommand{\PTA}{\pta}

\newcommand{\pedges}{\mathit{prob}}
\newcommand{\apedge}{p}
\newcommand{\pd}{\wp}

\newcommand{\locs}{L}
\newcommand{\locinit}{\bar{\aloc}}

\newcommand{\inv}{\mathit{inv}}

\newcommand{\aloc}{l}
\newcommand{\anevent}{\mathfrak{e}}
\newcommand{\exanevent}{\widehat{\mathfrak{e}}}

\newcommand{\g}{g}

\newcommand{\semMDP}[1]{\anMDP[{#1}]}

\newcommand{\bzero}{\mathbf{0}}

\newcommand{\vinz}{\sat}

\newcommand{\ccons}{\mathit{\Psi}}
\newcommand{\ccl}{\psi}

\newcommand{\reset}[2]{\mathsf{Reset}({#1},{#2})}
\newcommand{\resetnameonly}{\mathsf{Reset}}

\newcommand{\adelay}{\delta}

\newcommand{\fraction}[1]{\mathsf{frac}({#1})}
\newcommand{\floor}[1]{\lfloor{#1}\rfloor}

\newcommand{\set}[1]{\{{#1}\}}
\newcommand{\ldot}{\, . \,}

\newcommand{\adtmc}{\mathcal{C}}
\newcommand{\dstates}{S}
\newcommand{\dinit}{\bar{\astate}}
\newcommand{\dtrans}{\mathbf{P}}
\newcommand{\dtransof}[1]{\dtrans^{#1}}
\newcommand{\dpaths}[1]{\mathit{Paths}^{#1}}
\newcommand{\dfinpaths}[1]{\mathit{Paths}^{#1}_*}
\newcommand{\dprob}[1]{\mathrm{Pr}^{#1}}
\newcommand{\adtmcparaone}[1]{\adtmc^{#1}}

\newcommand{\apath}{\mathbf{\afinpath}}

\newcommand{\anmdp}{\mathcal{M}}
\newcommand{\mstates}{S}
\newcommand{\minit}{\bar{\astate}}
\newcommand{\mactions}{A}
\newcommand{\mtrans}{\Delta}
\newcommand{\noaction}{\bot}
\newcommand{\mactionsof}[1]{\mactions({#1})}

\newcommand{\mpaths}[1]{\mathit{Paths}^{#1}}
\newcommand{\mfinpaths}[1]{\mathit{Paths}^{#1}_*}
\newcommand{\afinpath}{r}

\newcommand{\apa}[1]{\mathbf{a}}

\newcommand{\targetset}{T}

\newcommand{\strats}[1]{\Sigma^{#1}}

\newcommand{\probfin}[1]{\mathrm{Pr}^{#1}_{*}}

\newcommand{\afinpathrepr}[2]{\hat{\afinpath}} 

\newcommand{\labels}{\mathit{Labels}}
\newcommand{\alab}{\mu}
\newcommand{\alabseq}{\pi}

\newcommand{\dtmclab}{\mathsf{c}}
\newcommand{\dtmclabfrom}[1]{\dtmclab^{#1}}

\newcommand{\abij}{f}

\newcommand{\labseqset}{\Xi}
\newcommand{\trace}[2]{\mathsf{trace}[{#1}]({#2})}
\newcommand{\dpathsfromlab}[2]{\dpaths{{#1}}({#2})}
\newcommand{\dpathsfromstrat}[2]{\dpaths{\adtmcparaone{#1}}({#2})}

\newcommand{\mdplab}{\mathsf{m}}
\newcommand{\mdplabof}[1]{\mdplab^{#1}}

\newcommand{\divpaths}{\mathsf{Divergent}}
\newcommand{\asstrats}[1]{\strats{#1}_\mathrm{a.s.}}
\newcommand{\surestrats}[1]{\strats{#1}_\mathrm{sure}}

\newcommand{\adrstrat}{\tilde{\astrat}}
\newcommand{\drstrats}[1]{\strats{#1}_\mathrm{dr}}

\newcommand{\valof}[1]{\mathsf{val}(#1)}
\newcommand{\vallast}[1]{\mathsf{vl}(#1)}

\newcommand{\prefof}[2]{{#1}\! \! \downharpoonright_{#2}}
\newcommand{\grdof}[1]{\mathsf{grdof}({#1})}
\newcommand{\distof}[1]{\mathsf{distof}({#1})}

\newcommand{\dtmcpathof}[2]{\downarrow^{#1}_{\adtmcparaone{#1}}\!\!\!({#2})}

\newcommand{\exPTA}{\widehat{\PTA}}
\newcommand{\exclock}{\mathfrak{z}}

\newcommand{\anexloc}{\widehat{\aloc}}
\newcommand{\ticklocs}{\widehat{\locs}}
\newcommand{\exinv}{\widehat{\inv}}
\newcommand{\expedges}{\widehat{\pedges}}

\newcommand{\cecs}{\mathsf{CECs}}
\newcommand{\acec}{\alpha}
\newcommand{\anothercec}{\beta}
\newcommand{\yancec}{\gamma}

\newcommand{\cecof}[1]{[{#1}]}
\newcommand{\tsucc}[2]{\mathit{tsucc}[{#1}]({#2})}

\newcommand{\regMDP}[1]{\mathfrak{M}[{#1}]}
\newcommand{\regions}{\mathfrak{Regs}}
\newcommand{\reginit}{(\locinit,\cecof{\bzero})}
\newcommand{\regactions}{\mathfrak{A}}
\newcommand{\regtrans}{\blacktriangle}

\newcommand{\regactionsof}[1]{\regactions({#1})}

\newcommand{\afinpathreg}{\mathfrak{r}}
\newcommand{\inregpath}{\in}
\newcommand{\afinpathregfrom}[1]{[{#1}]}

\newcommand{\dfinpathsfromregpath}[2]{\dpaths{{#1}}_*({#2})}

\newcommand{\choices}[3]{\mathsf{Choices}[{#1},{#2}]{#3}}
\newcommand{\adrdelay}[1]{\tilde{\adelay}_{#1}}
\newcommand{\adrfinpath}{\tilde{\afinpath}}
\newcommand{\adrpath}{\tilde{\apath}}

\newcommand{\adrval}{\tilde{\aval}}
\newcommand{\adrvalof}[2]{\adrval^{#1}_{#2}}
\newcommand{\avalof}[2]{\aval^{#1}_{#2}}

\newcommand{\aregstratof}[1]{[{#1}]}

\newcommand{\aninj}{\iota}

\newcommand{\adelayparatwo}[2]{\adelay^{#1}_{#2}}

\newcommand{\prfn}{\texttt{p}}
\newcommand{\infvisit}[1]{\mathrm{inf}({#1})}
\newcommand{\parity}[1]{\mathsf{Parity}({#1})}

\newcommand{\sure}{\mathsf{S}}

\newcommand{\objsat}[3]{{#1},{#2} \sat {#3}}
\newcommand{\probthresh}[2]{\mathbb{P}_{{#1}{#2}}}
\newcommand{\thresh}{\unrhd}

\newcommand{\mof}{\varphi}
\newcommand{\anobj}{\phi}

\newcommand{\objsatmdp}[2]{{#1} \sat {#2}}

\newcommand{\objsatpta}[2]{{#1} \sat {#2}}
\newcommand{\asobjsatpta}[2]{{#1} \sat_\mathrm{a.s.} {#2}}
\newcommand{\sureobjsatpta}[2]{{#1} \sat_\mathrm{sure} {#2}}

\newcommand{\exprfn}{\widehat{\prfn}}
\newcommand{\exmof}{\widehat{\mof}}

\newcommand{\tickio}{\square \Diamond \mathsf{tick}}
\newcommand{\tickprfn}{\prfn^\mathsf{tick}}

\newcommand{\ams}{\mathtt{m}}

\newcommand{\memstates}{\mathtt{M}}
\newcommand{\meminit}{\bar{\mathtt{m}}}
\newcommand{\memupd}{\mathtt{u}}
\newcommand{\nafn}{\mathtt{na}}

\newcommand{\extendmemupd}{\widehat{\memupd}}

\newcommand{\ptime}{\textsc{Ptime}}
\newcommand{\npconp}{\textsc{NP} \cap \textsc{coNP}}
\newcommand{\exptime}{\textsc{Exptime}}
\newcommand{\ncoexptime}{\textsc{Nexptime} \cap \textsc{coNexptime}}
\newcommand{\twoexptime}{\textsc{2Exptime}}

\begin{document}
%
\title{Probabilistic Timed Automata with\\Multiple Sure and Threshold Objectives}
\titlerunning{PTA with Multiple Sure and Threshold Objectives}
%
\author{Jeremy Sproston\orcidlink{0000-0003-2101-6821}}
\authorrunning{J. Sproston}
%
\institute{University of Turin, Italy\\
\email{jeremy.sproston@unito.it}}
\maketitle              
\begin{abstract}
Probabilistic timed automata extend the classical timed automata formalism with discrete probabilistic choice over edges.
We consider the case of the verification of a combination of multiple objectives of probabilistic timed automata,
where the objectives can be non-probabilistic (referring to sure satisfaction of all behaviors)
and probabilistic (referring to a probability threshold),
and refer to reachability or $\omega$-regular properties on executions expressed using parity conditions.
We show that existing algorithms for multiple objectives for finite-state Markov decision processes
can be applied,
using the classical region-graph construction,
to obtain algorithms for the corresponding objectives of probabilistic timed automata.
In particular, the verification of a probabilistic timed automata
against multiple probabilistic-threshold sink-reachability objectives 
subject to the almost-sure divergence of time is $\exptime$-complete,
while verification against the combination of a sure property, including sure divergence of time,
and a single probabilistic-threshold parity objective is in 
$\ncoexptime$.
%
%
\keywords{Probabilistic model checking  \and Timed automata \and Omega-regular
\and Multiple objectives.}
\end{abstract}

\section{Introduction}\label{sec:introduction}
During the development of complex systems,
the consideration of quantitative characteristics,
such as time, probability or energy,
can be of vital importance.
Formal methods for the description and automatic analysis of systems 
exhibiting such quantitative features 
have been developed accordingly.
In particular, the formalism of timed automata \cite{AD94} is well-established for
the description of timed systems.
A timed automaton comprises a finite directed graph equipped with a set of real-valued variables
called \emph{clocks}, which increase at the same rate as real-time,
and where the graph is labelled with constraints on clocks restricting the times at which 
its edges can be traversed.
Approaches for the incorporation of other quantitative information within 
the classical timed automaton formalism include those considering probabilistic behavior.
In this paper we concentrate on \emph{probabilistic timed automata} (PTAs)
\cite{GJ95,KNSS02,NPS13}.
PTAs extend conservatively timed automata by featuring,
in addition to the usual nondeterminsitic choice
over exact time delays and edges to be traversed,
the possibility of selecting which edges are traversed according to probabilistic choice.
PTAs can alternatively be seen as a conservative extension
of Markov decision processes (MDPs) with clocks and constraints on clocks.
The analysis of PTAs consists typically of the construction of 
an equivalent finite-state MDP \cite{KNPS06,KNSS02},
which is then subject to analysis methods for finite-state MDPs,
and has been implemented in tools such as {\sc Prism} \cite{KNP11}
and {\sc mcsta} \cite{HH14}.
PTAs and their associated analysis techniques 
have been used to model and analyze a number of case studies \cite{KNPS06,NPS13}.
PTAs can also be used as an abstract model for approximate analysis of 
stochastic systems with complex interplay between timing 
and stochastic characteristics \cite{HHH14}.
Analysis of PTAs has typically taken the form of the verification of 
whether the PTA satisfies a probabilistic version of
a \emph{single} reachability, temporal logic or $\omega$-regular property \cite{KNSS02,Spr11}
or a deterministic timed automaton \cite{FLL18}.
More precisely, such analysis aims at identifying whether there exists a \emph{strategy} 
(a way to resolve the nondeterministic choice of the PTA)
such that the probability that the strategy's behaviors satisfy a property is above a certain threshold. 

\begin{figure}[t]\label{fig:ptadiv}
\begin{center}
\resizebox{9cm}{!}{
\scriptsize
\begin{tikzpicture}[->,>=stealth',shorten >=1pt,auto, thin] 
  \tikzstyle{state}=[draw=black, text=black, shape=circle, inner sep=3pt, outer sep=0pt, circle, rounded corners] 
  \tikzstyle{final_state}=[draw=black, text=black, shape=circle, inner sep=3pt, outer sep=0pt, circle, rounded corners, fill= lg] 
  \tikzstyle{bad_state}=[draw=black, text=black, shape=circle, inner sep=3pt, outer sep=0pt, circle, rounded corners] 

	\node (for_initial){};
	\node[state, node distance=1.2cm](A) [below of=for_initial] {$\aloc_0$};
	\node[yshift=0.4cm,below of = A](inv_A){\colorbox{blue!7}{$\aclock \leq 1$}};
	\node[fill=black, node distance=2.1cm] (nail_A) [right of=A] {};
	\node[state, node distance=2.1cm](B)[right of=nail_A]{$\aloc_1$}; 
	\node[yshift=0.4cm,below of = B](inv_B){\colorbox{blue!7}{$\aclock \leq 1$}};

	\node[state, node distance=2.1cm](C)[left of=A]{$\aloc_2$}; 
	\node[yshift=0.4cm,below of = C](inv_C){\colorbox{blue!7}{$\aclock \leq 1$}};

	\path[rounded corners]
	(for_initial)
		edge (A)
	(A)
		edge [bend left] node[above] {\colorbox{blue!7}{$\anotherclock > 0$}} (nail_A)
		edge node[above] {\begin{tabular}{c}{\colorbox{blue!7}{$\aclock \leq 1$}}\\{\colorbox{red!15}{$1$}} \end{tabular}} (C)
	(nail_A)
		edge [above] node {\colorbox{red!15}{$\frac{1}{2}$}} (B)
		edge [bend left] node[below,text=black,pos=.1] {\colorbox{red!15}{$\frac{1}{2}$}}  (A)
		edge [bend left] node[below,text=black,pos=.5] {\colorbox{blue!7}{$\set{\anotherclock}$}} (A)
	(B)
		edge [loop above] node {\begin{tabular}{c}\colorbox{blue!7}{$\aclock = 1$}\\{\colorbox{red!15}{$1$}} \colorbox{blue!7}{$\set{\aclock}$}\end{tabular}} (B)
	(C)
		edge [loop above] node {\begin{tabular}{c}\colorbox{blue!7}{$\aclock = 1$}\\{\colorbox{red!15}{$1$}} \colorbox{blue!7}{$\set{\aclock}$}\end{tabular}} (C);

\end{tikzpicture}
}
\end{center}
\caption{A example of a PTA.}
\end{figure}
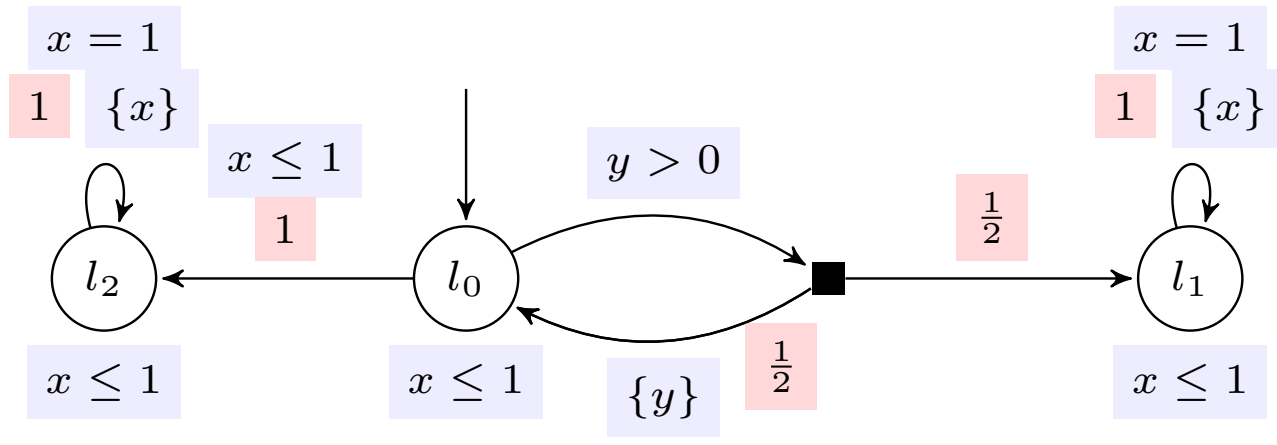

\begin{example}\label{ex:ptadiv}
An example of PTA is shown in \figref{fig:ptadiv}.
We use the usual conventions for drawing timed automata models,
and note that the clock constraints (e.g., $\aclock \leq 1$) 
and resets (e.g., $\set{\anotherclock}$) are shown with a blue background,
whereas probabilities are shown with a red background.
Consider location $\aloc_0$, and assume that the values of the clocks $\aclock$ and $\anotherclock$
are equal to $0$. 
A nondeterministic choice is made regarding the amount
of time that elapses in location $\aloc_0$ before exiting,
constrained by the fact that the value of $\aclock$ cannot exceed $1$
(by the invariant condition written below $\aloc_0$).
If a non-zero amount of time elapses in $\aloc_0$, 
then the value of $\anotherclock$ exceeds $0$ when $\aloc_0$ is exited,
and a nondeterministic choice is then made between the edge 
that leads to $\aloc_2$ with probability $1$,
and the edge which leads to the black square.
The latter edge bifurcates into two 
probabilistic alternatives,
one which leads to $\aloc_1$ with probability $\frac{1}{2}$,
and another which leads back to $\aloc_0$ with probability $\frac{1}{2}$,
in which case the clock $\anotherclock$ is reset to $0$
(as indicated by the edge label $\set{\anotherclock}$).
If $\anotherclock$ is $0$ when $\aloc_0$ is exited,
then the only available option is to take the edge to $\aloc_2$
(because the guard condition $\anotherclock > 0$ on the rightward edge from $\aloc_0$
is not satisfied).
Note that there exist strategies of the PTA to reach $\aloc_1$ with probability $1$,
and strategies to reach $\aloc_2$ with probability $1$.
\end{example}

Multi-objective threshold properties for MDPs specify that 
a \emph{combination of multiple properties}, each with their own threshold,
should be satisfied by a strategy of the MDP \cite{CMH06,EKVY08},
and have been used in a number of application contexts,
such as autonomous driving \cite{CKSW13} 
and task planning in mobile robotics \cite{LPH17}.
Methods for determining whether a multi-objective property is satisfied by an MDP
(and related problems, such as the approximation of the Pareto curve
corresponding to the multiple objectives under consideration)
have been implemented in {\sc Prism} \cite{FKNPQ11,FKP12},  
{\sc mcsta} and {\sc Storm} \cite{HQvW26}.

Furthermore, multi-objective properties have been extended to include
\emph{sure} objectives:
a strategy satisfies a sure objective 
if all behaviors of the strategy 
satisfies a property on behaviors.
In particular, the case of a multi-objective property combining 
a single sure objective and a single threshold parity objective,
and a single sure objective and multiple threshold sink-reachability objectives, 
have been studied in \cite{BRR17} and \cite{BKW24}, respectively
(where sink-reachability objectives are reachability objectives in which the target states 
are absorbing, i.e., once reached, we remain within the current state).

In this paper, we extend the basic single-objective approach for PTAs
to permit the verification of PTAs against multiple objectives.
Consider again the PTA of Example~\ref{ex:ptadiv}
and the multi-objective property specifying that 
$\aloc_1$ is reached with probability (at least) $\frac{1}{2}$
and that $\aloc_2$ is reached with probability (at least) $\frac{1}{2}$
(written typically using syntax such as 
$\probthresh{\geq}{\frac{1}{2}}(\Diamond \aloc_1) 
\wedge \probthresh{\geq}{\frac{1}{2}}(\Diamond \aloc_2)$).
This multi-objective property is satisfied by the PTA:
an example of a strategy that witnesses the satisfaction of the property 
is that which chooses the edge to the black square on the first visit to $\aloc_0$,
then chooses the edge to $\aloc_2$ if the PTA returns to $\aloc_0$.

Our approach for verifying multi-objective properties of PTA
is to consider the application of classical region equivalence \cite{AD94}
to obtain a finite-state MDP that is then subject to 
analysis methods for multiple-objective properties of finite-state MDPs.
We focus on the following multi-objective properties:
\begin{inparaenum}[(i)]
\item
multiple-threshold sink-reachability properties,
\item
properties with one sure parity objective and one threshold reachability or parity objective,
and 
\item
properties with one sure objective and multiple threshold sink-reachability objectives.
\end{inparaenum}
While we show that the standard region-graph MDP (see, for example, \cite{KNSS02}) 
can be used for verifying these multi-objective properties,
the proof of its correctness in the multiple-objective setting is complicated by the fact that 
randomization between nondeterministic choices is required for multiple-threshold objectives,
even in the finite-state MDP setting \cite{EKVY08}.
Previous work on PTAs has been limited mainly to non-randomized strategies,
which is sufficient in the case in which non-randomized strategies
capture optimal behavior on the finite-state region-graph MDP.
We show that randomization over time delays can be necessary for satisfying 
a multiple-objective threshold property in PTAs,
but also that we can limit attention to strategies for which time delays involved 
in a randomized choice lead to different regions,
without affecting the satisfaction of the multiple-objective property.
This latter property is key to showing the correctness of the region-graph MDP construction.
Given that the size of the region-graph MDP is exponential in the size of the description of the PTA,
and taking into account the complexity results for multi-objective properties on finite-state MDPs
of \cite{BKW24,BRR17,EKVY08},
we conclude that verifying PTA against 
\begin{inparaenum}[(i)]
\item
multiple-threshold sink-reachability objectives is 
$\exptime$-complete,
\item
one sure parity objective and one threshold reachability or parity objective
is in $\ncoexptime$,
and 
\item
one sure objective and multiple threshold sink-reachability objectives
is in $\twoexptime$.
\end{inparaenum}

Finally, we also consider issues of time divergence for PTAs.
In PTAs, as in timed automata,
behavior in which time does not elapse beyond any bound is unrealistic,
and should be excluded from consideration during analysis.
Previous work on divergence in PTAs has focused on \emph{almost-sure divergence},
which considers only strategies for which time diverges with probability $1$ \cite{KNSS02},
and \emph{sure divergence},
which considers only strategies for which time diverges on all behaviors \cite{Spr09}.
Note that, in Example~\ref{ex:ptadiv},
location $\aloc_1$ can be reached from $\aloc_0$ with probability $1$ under almost-sure strategies,
but cannot be reached with probability $1$ under sure strategies.
We show that almost-sure divergence can be represented in the multiple-objective framework
as an objective with a probability threshold of $1$,
and sure divergence can be represented as a sure objective.
This observation leads to new results for PTA.
First, the restriction to almost-sure time divergent strategies in the context of verifying a 
multiple-objective threshold property
can be handled by modifying the PTA (without any complexity blow up)
and adding a conjunct to the multiple-objective threshold property, 
meaning that we remain in the PTA/multiple-objective threshold property setting.
Second, we obtain a method for verifying a single-objective threshold parity objective,
and multiple-objective threshold sink-reachability objectives, 
in the context of 
sure divergence.
Third, by adapting a result of \cite{BRR17}, infinite memory is required in general
for a threshold coB\"{u}chi objective under sure divergence.

We note that strategies randomizing over edges are featured in
the work of \cite{JKPPS25,JKPPS26} on multi-agent systems
based on PTA.
Instead, the strategies of this paper make 
randomized choice over both edges and time delays.

\section{Preliminaries}\label{sec:preliminaries}

%
We use $\Rset$ to denote the set of real numbers,
$\nnr$ to denote the set of non-negative real numbers,
$\Qset$ to denote the set of rational numbers, and
$\Nset$ to denote the set of natural numbers.
%
%
%
A (discrete probability) \emph{distribution} over a countable set $Q$ is
a function ${\adist: Q \ra [0,1] \cap \Qset}$ such that $\sum_{q \in Q} \adist(q) = 1$.
Let $\dist(Q)$ be the set of distributions over $Q$.
For a (possibly uncountable) set $Q$ and a function $\adist : Q \ra [0,1] \cap \Qset$, 
we define $\support(\adist) = \set{q \in Q \setsep \adist(q) > 0}$.
For an uncountable set $Q$, we define $\dist(Q)$ to be the set of
functions ${\adist : Q \ra [0,1] \cap \Qset}$ such that $\support(\adist)$ is a
countable set and $\adist$ restricted to $\support(\adist)$ is a distribution.
Given a binary function $f: Q \times Q \ra [0,1]$ and $q \in Q$,
we denote by $f(q,\cdot): Q \ra [0,1]$ the unary function 
such that $f(q,\cdot)(q')=f(q,q')$ for each $q' \in Q$.

\subsubsection{Markov Chains.}
A \emph{Markov chain}
(MC) $\adtmc$ is a tuple $(\dstates,\dinit,\dtrans)$
where $\dstates$ is a set of states with initial state $\dinit \in \dstates$,
and $\dtrans: \dstates \times \dstates \ra [0,1]$ is a transition probability function
such that $\dtrans(\astate,\cdot) \in \dist(\dstates)$ for each state $\astate \in \dstates$.
An \emph{(infinite) path} of MC $\adtmc$ is an infinite sequence $\astate_0 \astate_1 \cdots$ of states
such that $\dtrans(\astate_i,\astate_{i+1})>0$ for all $i \geq 0$.
Given a path $\apath = \astate_0 \astate_1 \cdots$ and $i \geq 0$,
we let $\apath(i) = \astate_i$ be the $(i+1)$-th state along $\apath$.
The set of paths of $\adtmc$ starting in initial state $\dinit$ is denoted by $\dpaths{\adtmc}$.
Similarly,  a finite path of $\adtmc$ is a finite sequence $\afinpath = \astate_0 \astate_1 \cdots \astate_n$
such that  $\dtrans(\astate_i,\astate_{i+1})>0$ for all $0 \leq i < n$.
We let $\afinpath(i) = \astate_i$ for $0 \leq i \leq n$ be the $(i+1)$-th state along $\afinpath$,
and let $\dfinpaths{\adtmc}$ be the set of finite paths of $\adtmc$ starting in initial state $\dinit$.
Given a finite path $\afinpath = \astate_0 \astate_1 \cdots \astate_{n-1} \astate_n$ in $\dfinpaths{\adtmc}$,
let $\probfin{\adtmc}(\afinpath) 
= \dtrans(\astate_0,\astate_1) \cdot \ldots \cdot \dtrans(\astate_{n-1},\astate_n)$.
We can extend uniquely $\probfin{\adtmc}$ to obtain 
a probability measure $\dprob{\adtmc}$ over $\dpaths{\adtmc}$
(for more details, see \cite{BK08}).

Let $\labels$ be a finite set of \emph{labels},
and let $\dtmclab: \dstates \ra \labels$ 
be a function labeling states of $\adtmc$ with labels.
We refer to $(\adtmc,\dtmclab)$ as a \emph{labeled MC}.
For an infinite path $\apath = \astate_0 \astate_1 \cdots$ of $\adtmc$
let $\trace{\dtmclab}{\apath} = \dtmclab_1(\astate_0) \dtmclab_1(\astate_1) \cdots$
be its associated sequence of labels.
Let $\adtmc_1 = (\dstates_1,\dinit_1,\dtrans_1)$ and $\adtmc_2 = (\dstates_2,\dinit_2,\dtrans_2)$ be MCs,
and let $(\adtmc_1,\dtmclab_1)$ and $(\adtmc_2,\dtmclab_2)$
be labeled MCs.
Then $(\adtmc_1,\dtmclab_1)$ and $(\adtmc_2,\dtmclab_2)$ are \emph{isomorphic}
if there exists a bijection $\abij: \dstates_1 \ra \dstates_2$
such that
(1)~$\abij(\dinit_1) = \dinit_2$,
and (2)~for states $\astate_1 \in \dstates_1$ and $\astate_2 \in \dstates_2$,
if $\abij(\astate_1) = \astate_2$ then
(2a)~$\dtmclab_1(\astate_1) = \dtmclab_2(\astate_2)$,
and (2b)~for each state $\astate_1' \in \dstates_1$ 
we have $\dtrans_1(\astate_1,\astate_1') = \dtrans_2(\astate_2,\abij(\astate_1'))$.
\begin{lemma}\label{lem:isopath}
For isomorphic labeled MCs $(\adtmc_1,\dtmclab_1)$ and $(\adtmc_2,\dtmclab_2)$,
for a set $\labseqset \subseteq \labels^\omega$ of infinite sequences of labels,
we have that 
$\trace{\dtmclab_1}{\apath_1} \in \labseqset$ 
for all $\apath_1 \in \dpaths{\adtmc_1}$ 
if and only if 
$\trace{\dtmclab_2}{\apath_2} \in \labseqset$
for all $\apath_2 \in \dpaths{\adtmc_2}$.
\end{lemma}
Let $(\adtmc,\dtmclab)$ be a labeled MC,
and let $\labseqset \subseteq \labels^\omega$ be a set of sequences of labels.
If the set $\dpathsfromlab{\adtmc}{\labseqset}
= \set{ \apath \in \dpaths{\adtmc} \setsep \trace{\dtmclab}{\apath} \in \labseqset}$
is measurable, 
we say that $\labseqset$ is an 
\emph{event of $(\adtmc,\dtmclab)$}.
\begin{lemma}\label{lem:isoprob}
For event $\labseqset$ of isomorphic labeled MCs 
$(\adtmc_1,\dtmclab_1)$ and $(\adtmc_2,\dtmclab_2)$,
we have $\dprob{\adtmc_1}(\dpathsfromlab{\adtmc_1}{\labseqset})
= \dprob{\adtmc_2}(\dpathsfromlab{\adtmc_2}{\labseqset})$.
\end{lemma}
\lemref{lem:isoprob} follows from \cite[Lemma~10.66]{BK08}
(which shows a similar result for bisimulation,
a weaker equivalence than isomorphism).

\subsubsection{Markov Decision Processes.}
A \emph{Markov decision process} (abbreviated as MDP)
$\anmdp = ( \mstates, \minit, \mactions, \mtrans )$ comprises 
a set of states $\mstates$,
an initial state $\minit \in \mstates$,
a set of actions $\mactions$,
and a probabilistic transition function
$\mtrans: \mstates \times \mactions \ra \dist(\mstates) \cup \set{ \noaction }$.
The symbol $\noaction$ represents the unavailability of an action in a state,
i.e., $\mtrans(\astate,\anaction) = \noaction$ signifies that 
action $\anaction \in \mactions$ is not available in state $\astate \in \mstates$.
For each state $\astate \in \mstates$,
let $\mactionsof{\astate} = \set{ \anaction \in \mactions \setsep \mtrans(\astate,\anaction) \neq \noaction }$
and assume that $\mactionsof{\astate} \neq \emptyset$,
i.e., there is at least one available action in each state.
A transition from state $\astate$ is made according to two steps:
the first selects nondeterministically an action $\anaction$ from $\mactionsof{\astate}$,
while the second chooses probabilistically 
a target state according to the distribution $\mtrans(\astate,\anaction)$
(that is, target state $\astate'$ is chosen with probability $\mtrans(\astate,\anaction)(\astate')$). 
In general, the sets of states and actions
can be uncountable.
We say that an MDP is \emph{finite} if $\mstates$ and $\mactions$ are finite sets.
A(n infinite) path of an MDP $\anmdp$ is a sequence $\astate_0 \anaction_0 \astate_1 \anaction_1 \cdots$
such that $\anaction_i \in \mactionsof{\astate_i}$ and $\mtrans(\astate_i,\anaction_i)(\astate_{i+1})>0$
for all $i \geq 0$.
Given an infinite path $\apath = \astate_0 \anaction_0 \astate_1 \anaction_1 \cdots$ and $i \geq 0$,
we let $\apath(i) = \astate_i$ be the $(i+1)$-th state along $\apath$.
Let $\mpaths{\anmdp}$ be the set of infinite paths of $\anmdp$ starting in the initial state $\minit$.
A finite path is a sequence $\afinpath = \astate_0 \anaction_0 \astate_1 \anaction_1 \cdots \anaction_{n-1} \astate_n$
such that $\anaction_i \in \mactionsof{\astate_i}$ and $\mtrans(\astate_i,\anaction_i)(\astate_{i+1})>0$
for all $0 \leq i < n$.
Let $\last(\afinpath) = \astate_n$ denote the final state of $\afinpath$. 
For $\anaction \in \mactionsof{\astate_n}$ and $\astate \in \mstates$
such that $\mtrans(\astate_n,\anaction)(\astate)>0$,
we use $\afinpath \anaction \astate$ to denote the finite path
$\astate_0 \anaction_0 \astate_1 \anaction_1 \cdots \anaction_{n-1} \astate_n  \anaction \astate$.
Let $\mfinpaths{\anmdp}$ be the set of finite paths of the MDP $\anmdp$ starting in the initial state $\minit$.  

A \emph{strategy} of $\anmdp$ is a function 
$\astrat: \mfinpaths{\anmdp} \ra \bigcup_{\astate \in \mstates} \dist(\mactionsof{\astate})$
such that $\astrat(\afinpath) \in \dist(\mactionsof{\last(\afinpath)})$
and $|\support(\astrat(\afinrun))|$ is finite,
for all $\afinpath \in \mfinpaths{\anmdp}$.
Let $\strats{\anmdp}$ be the set of strategies of the MDP $\anmdp$.
We say that infinite path 
$\apath = \astate_0 \anaction_0 \astate_1 \anaction_1 \cdots$ is \emph{generated} by $\astrat$
if $\astrat(\astate_0 \anaction_0 \astate_1 \anaction_1 \cdots \anaction_{i-1} \astate_i)(\anaction_i)>0$
for all $i \in \Nset$.
Let $\mpaths{\astrat}$ be the set of paths generated by $\astrat$.
The set $\mfinpaths{\astrat}$ of finite paths generated by $\astrat$ is defined similarly.
Given a strategy $\astrat \in \strats{\anmdp}$,
we can define a countably infinite-state MC $\adtmcparaone{\astrat}$,
called the \emph{induced MC of $\astrat$},
that corresponds to the behaviour of $\astrat$:
we let $\adtmcparaone{\astrat} = (\mfinpaths{\astrat},\minit,\dtransof{\astrat})$,
where, for $\afinpath, \afinpath' \in \mfinpaths{\astrat}$,
we have $\dtransof{\astrat}(\afinpath, \afinpath') = \astrat(\afinpath)(\anaction) \cdot \mtrans(\last(\afinpath),\anaction)(\astate)$
if $\afinpath' = \afinpath \anaction \astate$ and $\anaction \in \mactionsof{\last(\afinpath)}$,
and $\dtransof{\astrat}(\afinpath, \afinpath') = 0$ otherwise.
Note that there is a 1-to-1 relationship between (finite and infinite) paths of $\anmdp$ generated by $\astrat$
and paths of $\adtmcparaone{\astrat}$:
for example, infinite path 
$\astate_0 \anaction_0 \astate_1 \anaction_1 \cdots$ of $\mpaths{\astrat}$
corresponds to the infinite path 
$\astate_0 (\astate_0 \anaction_0 \astate_1) (\astate_0 \anaction_0 \astate_1 \anaction_1 \astate_2) \cdots$
of $\dpaths{\adtmcparaone{\astrat}}$.
This allows us to define a probability measure, denoted by $\dprob{\astrat}$, 
over $\mpaths{\astrat}$
from the probability measure $\dprob{\adtmcparaone{\astrat}}$ over $\dpaths{\adtmcparaone{\astrat}}$.

Given a finite set $\labels$ of labels,
let $\mdplab: \mstates \ra \labels$ 
be a labeling function for $\anmdp$,
and refer to $(\anmdp,\mdplab)$ 
as a \emph{labeled MDP}.
Let $\astrat \in \strats{\anmdp}$ be a strategy of $\anmdp$.
We define the \emph{labeled MC induced by $\astrat$ and $\mdplab$}
as $(\adtmcparaone{\astrat},\dtmclabfrom{\mdplab})$
where $\dtmclabfrom{\mdplab}(\afinpath) = \mdplab(\last(\afinpath))$
for each $\afinpath \in \mfinpaths{\astrat}$.

Next, we extend the notions of traces and events to labeled MDPs.
Let $(\anmdp,\mdplab)$ be a labeled MDP.
For an infinite path $\apath = \astate_0 \astate_1 \cdots$ of $\anmdp$
let $\trace{\mdplab}{\apath} = \mdplab(\astate_0) \mdplab(\astate_1) \cdots$.
We say that a set $\labseqset \subseteq \labels^\omega$ is an event
if,  for each $\astrat \in \strats{\anmdp}$, 
the sets $\dpathsfromstrat{\astrat}{\labseqset}
= \set{ \apath \in \dpaths{\adtmcparaone{\astrat}} \setsep \trace{\dtmclabfrom{\mdplab}}{\apath} \in \labseqset}$
is measurable.

In the sequel, we consider two kinds of events:
reachability properties and parity conditions.
A \emph{label sequence} $\alabseq = \alab_0 \alab_1 \alab_2 \cdots$
is an infinite sequence of labels (that is, $\alabseq \in \labels^\omega$).
The $(i+1)$-th label along $\alabseq$ is denoted by $\alabseq(i)$, i.e., $\alabseq(i) = \alab_i$.
Let $\targetset \subseteq \labels$ be a set of target labels.
We write the \emph{reachability property} $\Diamond \targetset$
to denote the set of infinite sequences of labels including at least one label in $\targetset$.
Formally 
$\Diamond \targetset
= \set{ \alabseq \in \labels^\omega \setsep \exists i \in \Nset \ldot \alabseq(i) \in \targetset  }$.
A \emph{priority function} $\prfn: \labels \ra \set{ 1, 2, \ldots, d }$ assigns a priority to each label.
Given a label sequence $\alabseq \in \labels^\omega$,
let $\infvisit{\alabseq} = \set{  k \setsep \forall i \geq 0 . \exists j \geq i . \prfn(\alabseq(j)) = k }$
be the set of priorities observed infinitely often along $\alabseq$.
The \emph{parity condition} associated with $\prfn$ corresponds to label sequences along which 
the maximum priority over those priorities that are observed infinitely often is even.
Formally, the set of label sequences of the parity condition associated with $\prfn$ 
is defined as 
$\parity{\prfn} = 
\set{ \alabseq \in \labels^\omega \setsep \max\set{k \in \infvisit{\alabseq}} \mbox{ is even} }$.
Henceforth, we assume that we consider only labeled MDPs 
for which reachability properties and parity conditions are measurable.
A reachability property $\Diamond \targetset$ of an labeled MDP 
is a \emph{sink-reachability property}
if each state $\astate \in \mstates$ such that $\mdplab(\astate) \in \targetset$
is absorbing, i.e.,
each available action $\anaction \in \mactionsof{\astate}$ is such that 
$\mtrans(\astate,\anaction)(\astate) = 1$.

A \emph{sure}  \emph{objective}
based on event $\anevent \subseteq \labels^\omega$,
written $\sure(\anevent)$ is satisfied by a strategy of $\anmdp$
if all of the strategy's traces are in $\anevent$.
Formally, given $(\anmdp,\mdplab)$ and $\astrat \in \strats{\anmdp}$, 
we write $\objsat{(\anmdp,\mdplab)}{\astrat}{\sure(\anevent)}$
if and only if $\trace{\mdplab}{\apath} \in \anevent$ for all $\apath \in \mpaths{\astrat}$.
A \emph{threshold objective} based on event $\anevent$ of $\anmdp$,
with $\thresh \in \set{ >, \geq }$ and $\lambda \in [0,1]$,
written $\probthresh{\thresh}{\lambda}(\anevent)$,
is satisfied by a strategy of $\anmdp$
if the probability of its traces that are in $\anevent$
is related by $\thresh$ to $\lambda$.
Formally, given $(\anmdp,\mdplab)$ and $\astrat \in \strats{\anmdp}$,
we write $\objsat{(\anmdp,\mdplab)}{\astrat}{\probthresh{\thresh}{\lambda}(\anevent)}$
if and only if $\dprob{\astrat}(\dpathsfromstrat{\astrat}{\anevent}) \thresh \lambda$.

A \emph{multi-objective formula} (with respect to $\labels$) consists of a conjunction of objectives
based on events,
that is $\mof = \bigwedge_{i=1}^n \anobj_i(\anevent_i)$,
where 
$\anobj_i$ is $\sure$ 
or belongs to  
$\set{\probthresh{\thresh}{\lambda} \setsep \thresh \in \set{ >, \geq } \mbox{ and } \lambda \in [0,1] }$,
and $\anevent_i$ is either a reachability property or a parity condition, for each $1 \leq i \leq n$.
Let $(\anmdp,\mdplab)$ be a labeled MDP.
Given strategy $\astrat \in \strats{\anmdp}$,
we write $\objsat{(\anmdp,\mdplab)}{\astrat}{\mof}$ if and only if
$\bigwedge_{i=1}^n \objsat{(\anmdp,\mdplab)}{\astrat}{\anobj_i(\anevent_i)}$,
and $\objsatmdp{(\anmdp,\mdplab)}{\mof}$ if and only if there exists $\astrat \in \strats{\anmdp}$
such that $\objsat{(\anmdp,\mdplab)}{\astrat}{\mof}$.
The following lemma is a consequence of \lemref{lem:isopath}
and \lemref{lem:isoprob}.
Its proof can be found in the appendix.

\begin{lemma}\label{lem:equiv_mdp_mof}
Let $(\anmdp_1,\mdplab_1)$ and 
$(\anmdp_2,\mdplab_2)$ be labeled MDPs,
such that, for all strategies $\astrat_1 \in \strats{\anmdp_1}$ of $\anmdp_1$,
there exists a strategy $\astrat_2 \in \strats{\anmdp_2}$  of $\anmdp_2$
for which 
$(\adtmcparaone{\astrat_1},\dtmclabfrom{\mdplab_1})$
and $(\adtmcparaone{\astrat_2},\dtmclabfrom{\mdplab_2})$ are isomorphic.
Let $\mof$ be a multi-objective formula.
If $\objsatmdp{(\anmdp_1,\mdplab_1)}{\mof}$
then $\objsatmdp{(\anmdp_2,\mdplab_2)}{\mof}$.
\end{lemma}

\section{Probabilistic Timed Automata}\label{sec:pta}
In this section, we recall the definition of probabilistic timed automata.
Let $\clocks$ be a finite set of \emph{clocks},
which are real-valued variables, the values of which
increase at the same rate as real-time.
We refer to
a mapping $\aval: \clocks \ra \nnr$ as a \emph{clock valuation}.
Let $\nnr^\clocks$ denote the set of clock valuations.
Let $\bzero \in \nnr^\clocks$ be the clock valuation which assigns 0 to all
clocks in $\clocks$.
For a clock valuation $\aval \in \nnr^\clocks$ and $\adelay \in \nnr$
(intended to represent a time delay), 
we use $\aval + \adelay$ to denote the clock valuation obtained by letting 
$(\aval + \adelay)(\aclock) = \aval(\aclock) + \adelay$ for all clocks $\aclock \in \clocks$. 
For a clock set $\aclockset \subseteq \clocks$, we let $\aval[\aclockset:=0]$ be the clock valuation
obtained from $\aval$ by resetting all clocks in $\aclockset$ to $0$;
formally, let $\aval[\aclockset:=0](\aclock)=0$ for all $\aclock \in \aclockset$,
and let $\aval[\aclockset:=0](\aclock)=\aval(\aclock)$ for all $\aclock \in \clocks \setminus \aclockset$.
For two clock valuations $\aval, \aval' \in \nnr^\clocks$,
let $\reset{\aval}{\aval'} = \set{ \aclockset \subseteq \clocks \mid \aval[\aclockset:=0] = \aval' }$
comprise the clock sets $\aclockset$ such that,
after resetting clocks in $\aclockset$ to $0$ in $\aval$,
the clock valuation $\aval'$ is obtained.
The set $\ccons$ of {\em clock constraints} over $\clocks$ is defined
as the set of conjunctions over atomic formulae of the
form
$\aclock \sim c$, where $\aclock,\anotherclock \in \clocks$,
$\sim \in \{ <,\leq,>,\geq,= \}$, and $c \in \Nset$.
A clock valuation $\aval \in \nnr^\clocks$ \emph{satisfies} the clock constraint $\ccl \in \ccons$, 
written $\aval \vinz \ccl$, if and only if $\ccl$ is satisfied 
after substituting each clock $\aclock \in \clocks$ in $\ccl$ with its value $\aval(\aclock)$.

A \emph{probabilistic timed automaton} (PTA)  $\PTA = (\locs, \locinit, \clocks, \inv, \pedges)$
comprises:
a finite set
$\locs$ of \emph{locations} with an \emph{initial location} $\locinit \in \locs$;
a finite set $\clocks$ of clocks;
a function $\inv: \locs \ra \ccons$ associating an \emph{invariant condition} with each location;
a set $\pedges \subseteq \locs \times \ccons \times \dist(2^{\clocks} \times \locs)$
of \emph{probabilistic edges}.
A probabilistic edge $(\aloc,\g,\pd) \in \pedges$ is a triple where:
\begin{inparaenum}[(1)]
\item $\aloc$ is the source location;
\item $\g$ is a clock constraint, called a \emph{guard}; and
\item $\pd$ is a probability distribution over $2^\clocks \times \locs$,
 i.e., over
  pairs $(\aclockset,\aloc')$ for clock set
  $\aclockset \subseteq \clocks$ and target location $\aloc' \in \locs$.
\end{inparaenum}

A \emph{state\/} of a PTA is a pair $(\aloc,\aval) \in \locs \times \valuations$ such that $\aval \vinz \inv(\aloc)$.
A transition of a PTA takes the following form.
From state $(\aloc,\aval)$, a time delay $\adelay \in \nnr$ 
is chosen nondeterministically,
subject to the constraint that 
the invariant $\inv(\aloc)$ must remain continuously satisfied
as $\adelay$ time elapses from $(\aloc,\aval)$.
Then, from the resulting state $(\aloc,\aval{+}\adelay)$, 
a probabilistic edge $(\aloc',\g,\pd) \in \pedges$ can be chosen nondeterministically
if $\aloc = \aloc'$ and 
its guard $\g$ is satisfied by $\aval{+}\adelay$.
After a probabilistic edge $(\aloc',\g,\pd)$ has been chosen, a set of clocks to reset and a successor location are selected at random,
according to the distribution $\pd$,
i.e., clocks in $\aclockset \subseteq \clocks$ are reset to $0$ 
and the PTA moves to location $\aloc''$
with probability $\pd(\aclockset,\aloc'')$,
resulting in the state $(\aloc'',(\aval{+}\adelay)[\aclockset:=0])$.

As is standard in the literature on PTAs (see, for example, \cite{NPS13}), 
the size of a PTA includes the binary encoding of both the constants used in the clock constraints 
(invariant conditions and guards),
and the encoding of probabilities as ratios between two natural numbers,
each written in binary.
%


Before presenting formally the semantics of PTAs,
we make the following standard well-formedness assumptions:
\begin{inparaenum}[(1)]
\item 
for any state $(\aloc,\aval)$, 
there exists $\adelay \in \nnr$ and probabilistic edge $(\aloc,\g,\pd) \in \pedges$
such that $\aval{+}\adelay \vinz \g$ 
and $\aval{+}\adelay' \vinz \inv(\aloc)$ for all $0 \leq \adelay' \leq \adelay$
(from any state, some transition is possible),
and 
\item
for any probabilistic edge $(\aloc,\g,\pd) \in \pedges$ 
and any clock valuation $\aval \in \nnr^\clocks$ such that $\aval \vinz \g \wedge \inv(\aloc)$,
we have $\aval[\aclockset:=0] \vinz \inv(\aloc')$ for each $(\aclockset,\aloc') \in \support(\pd)$
(probabilistic edges must lead to valid states).
\end{inparaenum}

The \emph{semantic MDP of the PTA
$\PTA = (\locs, \locinit, \clocks, \inv, \pedges)$}
is the MDP
$\semMDP{\PTA} = ( \mstates, \sinit, \mactions, \mdptrans )$
where
\begin{itemize}
\item
$\mstates = \{ (\aloc,\aval) \mid \aloc \in \locs \mbox{ and } \aval \in \nnr^\clocks \mbox{ s.t. } \aval \vinz \inv(\aloc) \}$
and $\sinit = (\locinit,\bzero)$;
\item
$\mactions = \nnr \times \pedges$;
\item
for $(\aloc,\aval) \in \mstates$, $\adelay \in \nnr$ and $(\aloc,\g,\pd) \in \pedges$
such that 
	\begin{inparaenum}[(i)]
	\item $\aval{+}\adelay \vinz \g$ and 
	\item $\aval{+}\adelay' \vinz \inv(\aloc)$ for all $0 \leq \adelay' \leq \adelay$,
	\end{inparaenum}
let $\mdptrans((\aloc,\aval),(\adelay,(\aloc,\g,\pd)))$ be the distribution such that,
for $(\aloc',\aval') \in \mstates$:
\begin{equation*}
\mdptrans((\aloc,\aval),(\adelay,(\aloc,\g,\pd)))(\aloc',\aval') 
=  
\sum_{\aclockset \in \reset{\aval{+}\adelay}{\aval'}} \pd(\aclockset,\aloc') \, ;
\end{equation*}
for $\adelay \in \nnr$ and $\apedge \in \pedges$ 
not satisfying conditions $\mathrm{(i)}$ and $\mathrm{(ii)}$,
we let $\mdptrans((\aloc,\aval),(\adelay,\apedge)) = \bot$.
\end{itemize}

We also consider a labeled extension of the semantic MDP,
where the labeling function labels a state with its corresponding location.
Formally, for PTA $\PTA$,
the \emph{labeled semantic MDP of $\PTA$} 
is the labeled MDP $(\semMDP{\PTA},\mdplabof{\semMDP{\PTA}})$,
where $\mdplabof{\semMDP{\PTA}}: \mstates \ra \locs$ is such that
$\mdplabof{\semMDP{\PTA}}(\aloc,\aval) = \aloc$ for each $(\aloc,\aval) \in \mstates$.
Hence, the events that we consider in the context of PTA
are reachability properties and parity conditions based on locations:
more precisely, given that $\labels = \locs$,
reachability properties take the form of $\Diamond \targetset$ for $\targetset \subseteq \locs$;
similarly, priority functions take the form of $\prfn: \locs \ra \set{ 1, 2, \ldots, d }$.
Given a multi-objective formula $\mof$ with respect to $\locs$,
we write $\objsatpta{\PTA}{\mof}$
if and only if $\objsatmdp{(\semMDP{\PTA},\mdplabof{\semMDP{\PTA}})}{\mof}$.

\section{Verifying Multi-objective Formulae of PTAs}\label{sec:correctness}

In the first part of this section,
we recall the definitions of region equivalence \cite{AD94}
and the region-graph MDP of a PTA \cite{KNSS02}.
We then 
consider properties of the strategies that witness the satisfaction of multi-objective formulae,
justify the correctness of the use of the region-graph MDP in this context,
and finally state the complexity of the verification problem of some classes of multi-objective formula.
Proofs of the results of this section can be found in the appendix.

\subsubsection{Region-graph MDP.}
For $t \in \nnr$,
we let $\fraction{t} = t - \floor{t}$.
For each clock $\aclock \in \clocks$ of a PTA $\PTA = (\locs, \locinit, \clocks, \inv, \pedges)$,
we let $c_\aclock$ be the maximal constant to which $\aclock$ is compared in any of the invariants or guards of probabilistic edges of $\PTA$
(if $\aclock$ is not involved in any clock constraint of $\PTA$, we let $c_\aclock=1$).
Two clock valuations $\aval,\aval' \in \nnr^\clocks$
are \emph{clock equivalent} if the following conditions are satisfied:
(1) for all clocks $\aclock \in \clocks$,
we have $\aval(\aclock) \leq c_\aclock$ if and only if $\aval'(\aclock) \leq c_\aclock$;
(2) for all clocks $\aclock \in \clocks$ with $\aval(\aclock) \leq c_\aclock$,
we have $\floor{\aval(\aclock)} = \floor{\aval'(\aclock)}$;
(3) for all clocks $\aclock,\anotherclock \in \clocks$ 
with $\aval(\aclock) \leq c_\aclock$ and $\aval(\anotherclock) \leq c_\anotherclock$,
we have $\fraction{\aval(\aclock)} \leq \fraction{\aval(\anotherclock)}$ 
if and only if 
$\fraction{\aval'(\aclock)} \leq \fraction{\aval'(\anotherclock)}$;
and (4) for all clocks $\aclock \in \clocks$ with $\aval(\aclock) \leq c_\aclock$,
we have $\fraction{\aval(\aclock)}=0$ if and only if $\fraction{\aval'(\aclock)}=0$.
Let $\cecs$ be the set of clock equivalence classes of $\PTA$.

Given a clock valuation $\aval \in \nnr^\clocks$, 
the unique clock equivalence class that contains $\aval$ is denoted by $\cecof{\aval}$.
Given a clock equivalence class $\acec \in \cecs$ and a clock constraint $\ccl$
that is either an invariant or a guard of $\PTA$,
we write $\acec \vinz \ccl$ to denote that 
there exists a clock valuation $\aval \in \acec$ such that $\aval \vinz \ccl$
(note also that the existence of $\aval \in \acec$ such that $\aval \vinz \ccl$
implies that $\aval' \vinz \ccl$ for all $\aval' \in \acec$).
We write $\acec[\aclockset:=0]$ to denote the set 
$\set{\aval[\aclockset:=0] \setsep \aval \in \acec}$.
Note that $\acec[\aclockset:=0]$ is itself a clock equivalence class.
We use 
$\reset{\acec}{\anothercec} = 
\set{ \aclockset \subseteq \clocks \mid \acec[\aclockset:=0] = \anothercec }$
to denote the set of clock sets that, when reset to $0$ from clock valuations in $\acec$,
result in $\anothercec$.
The subsequent \lemref{lem:reset_cor} relates $\resetnameonly$ for clock valuations
and for regions
(see \cite{Spr21FI} for a formal proof).

\begin{lemma}\label{lem:reset_cor}
For clock valuations $\aval, \aval' \in \nnr^\clocks$,
we have $\reset{\aval}{\aval'} = \reset{\cecof{\aval}}{\cecof{\aval'}}$.
\end{lemma}

Two states $(\aloc,\aval),(\aloc',\aval')$ are \emph{region equivalent} if
(1) $\aloc = \aloc'$, and (2) $\aval$ and $\aval'$ are clock equivalent.
A \emph{region} is an equivalence class of region equivalence.
Regions will be written as pairs $(\aloc,\acec)$, 
where $\aloc\in \locs$ is a location and $\acec \in \cecs$ is a clock equivalence class.
Let $\regions$ be the set of regions of $\PTA$.
Following standard results on the number of regions of timed automata \cite{AD94},
the number of regions corresponding to the PTA $\PTA$ is bounded by
$|\locs| \cdot \prod_{\aclock \in \clocks} (c_\aclock+1) \cdot
|\clocks| ! \cdot 2^{|\clocks|}$.

Let $(\aloc,\acec)$ be a region.
A \emph{time sucessor} of $(\aloc,\acec)$ 
is a clock equivalence class that can be reached from clock valuations in $\acec$
by letting time pass, subject to the invariant of $\aloc$:
formally $\anothercec \in \cecs$ is a time successor of $(\aloc,\acec)$
if there exist $\aval \in \acec$ and $\adelay \in \nnr$
such that $\aval{+}\adelay \in \anothercec$ and 
$\aval{+}\adelay' \vinz \inv(\aloc)$ for all $0 \leq \adelay' \leq \adelay$.
Let $\tsucc{\aloc}{\acec}$ be the set of time sucessors of $(\aloc,\acec)$.

The set of regions of PTA $\PTA$
can be used to construct the region-graph MDP of $\PTA$,
denoted by $\regMDP{\PTA} = (\regions, \reginit, \regactions, \regtrans)$, in the following way.
The set of states of $\regMDP{\PTA}$ is the set $\regions$ of regions,
and the initial state is $\reginit$.
The set of actions of $\regMDP{\PTA}$ is defined as 
$\regactions = \cecs \times \pedges$.
The probabilistic transition function 
$\regtrans: \regions \times \regactions \ra \dist(\regions)$ 
is defined as follows.
For $(\aloc,\acec) \in \regions$ and $(\anothercec,(\aloc',\g,\pd)) \in \regactions$
such that 
\begin{inparaenum}[(i)]
\item 
$\aloc = \aloc'$,
\item
$\anothercec \in \tsucc{\aloc}{\acec}$ and 
\item
$\anothercec \vinz \g$,
\end{inparaenum}
let $\regtrans((\aloc,\acec),(\anothercec,(\aloc',\g,\pd)))$ be the distribution
such that, for $(\aloc'',\yancec) \in \regions$:
\begin{equation*}
\regtrans((\aloc,\acec),(\anothercec,(\aloc',\g,\pd)))(\aloc'',\yancec)
=
\sum_{\aclockset \in \reset{\anothercec}{\yancec}} \pd(\aclockset,\aloc'') \, .
\end{equation*}
For $(\aloc,\acec) \in \regions$ and $(\anothercec,(\aloc',\g,\pd)) \in \regactions$
not satisfying at least one of conditions (i), (ii) or (iii),
let $\regtrans((\aloc,\acec),(\anothercec,(\aloc',\g,\pd))) = \noaction$.
Note that $\regMDP{\PTA}$ is a finite MDP.

As in the case of the labeled semantic MDP of a PTA,
we also consider the location of a region as its label:
formally, for PTA we let $\mdplabof{\regMDP{\PTA}}: \regions \ra \locs$
be such that $\mdplabof{\regMDP{\PTA}}(\aloc,\acec) = \aloc$
for each $(\aloc,\acec) \in \regions$.
The labeled region-graph MDP of $\PTA$ is 
the labeled MDP $(\regMDP{\PTA},\mdplabof{\regMDP{\PTA}})$.

In the sequel, we will show the following result,
which specifies that determining whether a PTA satisfies a multi-objective formula
can be achieved by determining whether the region-graph MDP satisfies the same formula.

\begin{proposition}\label{prop:mof_rg}
Let $\PTA$ be a PTA  
and $\mof$ be a multi-objective formula with respect to the set of locations of $\PTA$.
Then $\objsatpta{\PTA}{\mof}$ 
if and only if 
$\objsatmdp{(\regMDP{\PTA},\mdplabof{\regMDP{\PTA}})}{\mof}$.
\end{proposition}

\subsubsection{Region strategies, randomization and memory.}
Before establishing \propref{prop:mof_rg},
we introduce pure strategies
and region strategies (without memory and with finite-memory),
and consider their role in the multi-objective context.

A strategy $\astrat \in \strats{\semMDP{\PTA}}$ 
is \emph{pure} if $|\support(\astrat(\afinpath))| = 1$
for all finite paths $\afinpath \in \mfinpaths{\semMDP{\PTA}}$,
i.e., it does not choose randomly between its actions.

We now consider region strategies,
following the precedent of \cite{dAFHMS03,CHP08} 
for 2-player timed automata games.
We henceforth refer to finite paths of $\regMDP{\PTA}$
as \emph{finite region paths}.
Consider the following finite region path $\afinpathreg$ 
and finite path $\afinpath$ of $\semMDP{\PTA}$:
\begin{eqnarray*}
\afinpathreg & = & (\aloc_0,\acec_0)(\anothercec_0,\apedge_0)
(\aloc_1,\acec_1)(\anothercec_1,\apedge_1)
\ldots
(\anothercec_{n-1},\apedge_{n-1})
(\aloc_n,\acec_n)
\\
\afinpath & = & (\aloc_0',\aval_0)(\adelay_0,\apedge_0')
(\aloc_1',\aval_1)(\adelay_1,\apedge_1')
\ldots
(\adelay_{m-1},\apedge_{m-1}')
(\aloc_m',\aval_m) \; .
\end{eqnarray*}
We write $\afinpath \inregpath \afinpathreg$
if (1)~$n=m$, 
(2)~$\aloc_i = \aloc_i'$ and $\aval_i \in \acec_i$ for each $0 \leq i \leq n$,
and (3)~$\aval_i{+}\adelay_i \in \anothercec_i$ and $\apedge_i = \apedge_i'$
for each $0 \leq i < n$.
Let $\afinpathregfrom{\afinpath}$ be the unique finite region path of $\regMDP{\PTA}$ such that
$\afinpath \inregpath \afinpathregfrom{\afinpath}$
(given that $\regMDP{\PTA}$ represents all possible behaviors of $\PTA$,
such $\afinpathregfrom{\afinpath}$ exists for any $\afinpath \in \mfinpaths{\semMDP{\PTA}}$).

Given a state $\astate \in \mstates$,
we use $\valof{\astate}$ to refer to the clock valuation of $\astate$,
i.e., if $\astate = (\aloc,\aval)$ then $\valof{\astate} = \aval$.
Let $\astrat \in \strats{\semMDP{\PTA}}$ be a strategy
and $\afinpath,\afinpath' \in \mfinpaths{\semMDP{\PTA}}$ be finite paths.
We say that \emph{$\astrat$ has clock-equivalent choices for $\afinpath,\afinpath'$}
if 
there exists a bijection $\abij: \support(\astrat(\afinpath)) \ra \support(\astrat(\afinpath'))$
for which, for each $(\adelay,\apedge) \in \support(\astrat(\afinpath))$,
we have 
\begin{inparaenum}[(1)]
\item
$\abij(\adelay,\apedge)=(\adelay',\apedge)$
where $\cecof{\valof{\last(\afinpath)}{+}\adelay} = \cecof{\valof{\last(\afinpath')}{+}\adelay'}$,
and 
\item
$\astrat(\afinpath)(\adelay,\apedge) = \astrat(\afinpath')(\abij(\adelay,\apedge))$.
\end{inparaenum}

A strategy $\astrat \in \strats{\semMDP{\PTA}}$
is a \emph{region strategy} if it has clock-equivalent choices
for all finite paths $\afinpath,\afinpath' \in \mfinpaths{\semMDP{\PTA}}$
such that $\afinpathregfrom{\afinpath} = \afinpathregfrom{\afinpath'}$.
A region strategy $\astrat \in \strats{\semMDP{\PTA}}$ 
is \emph{memoryless} if it has clock-equivalent choices
for all finite paths $\afinpath,\afinpath' \in \mfinpaths{\semMDP{\PTA}}$ 
such that $\last(\afinpath)$ and $\last(\afinpath')$ are region equivalent.

A \emph{stochastic Moore machine} $(\memstates,\meminit,\memupd,\nafn)$
comprises a finite set $\memstates$ of memory states,
an initial memory state $\meminit \in \memstates$,
an update function $\memupd: \memstates \times \regions \ra \memstates$
and a next-action function $\nafn: \memstates \times \regions \ra \dist(\regactions)$.
Given a current region $(\aloc,\acec)$ and memory state $\ams \in \memstates$,
the stochastic Moore machine chooses the next action 
$(\anothercec,\apedge) \in \regactionsof{\aloc,\acec}$ with probability 
$\nafn(\ams,(\aloc,\acec))(\anothercec,\apedge)$,
and updates its memory state to $\memupd(\ams,(\aloc,\acec))$.
Given a finite region path $\afinpathreg \in \mfinpaths{\regMDP{\PTA}}$,
we let $\extendmemupd(\meminit,\afinpathreg)$ be 
the standard extension of $\memupd$ to sequences of regions along $\afinpathreg$,
i.e., $\extendmemupd(\meminit,\afinpathreg (\anothercec,\apedge)(\aloc,\acec))
= \memupd(\extendmemupd(\meminit,\afinpathreg),(\aloc,\acec))$
and $\extendmemupd(\meminit,(\aloc,\acec)) = \memupd(\meminit,(\aloc,\acec))$.
A region strategy $\astrat \in \strats{\semMDP{\PTA}}$
is a \emph{finite-memory region strategy} if it is defined according to 
a stochastic Moore machine $(\memstates,\meminit,\memupd,\nafn)$ in the following way.
For all finite paths $\afinpath,\afinpath' \in \mfinpaths{\semMDP{\PTA}}$
such that $\extendmemupd(\meminit,\afinpathregfrom{\afinpath}) 
= \extendmemupd(\meminit,\afinpathregfrom{\afinpath'})$:
\begin{inparaenum}[(1)]
\item
there exists a bijection 
$\abij: \support(\astrat(\afinpath)) \ra 
\support(\nafn(\extendmemupd(\meminit,\afinpathregfrom{\afinpath})))$
such that, for each $(\adelay,\apedge) \in \support(\astrat(\afinpath))$,
we have $\abij(\adelay,\apedge) = (\cecof{\valof{\last(\afinpath)}{+}\adelay},\apedge)$
and $\astrat(\afinpath)(\adelay,\apedge) = 
\nafn(\extendmemupd(\meminit,\afinpathregfrom{\afinpath}))(\abij(\adelay,\apedge))$,
and
\item
there exists a bijection
$\abij': \support(\astrat(\afinpath')) \ra 
\support(\nafn(\extendmemupd(\meminit,\afinpathregfrom{\afinpath'})))$
such that, for each $(\adelay',\apedge') \in \support(\astrat(\afinpath'))$,
we have $\abij'(\adelay',\apedge') = (\cecof{\valof{\last(\afinpath')}{+}\adelay'},\apedge')$
and $\astrat(\afinpath')(\adelay',\apedge') = 
\nafn(\extendmemupd(\meminit,\afinpathregfrom{\afinpath'}))(\abij'(\adelay',\apedge'))$.
\end{inparaenum}

Note that region strategies are of particular importance 
given the use of the region-graph MDP as a finite-state MDP that, 
from the perspective of the multi-objective formula that we consider,
is equivalent to the semantic MDP of the PTA.
In particular, the sufficiency of a certain set of strategies on the finite-state region-graph MDP
for a particular multi-objective formula 
implies the sufficiency of a related set of region strategies on the semantic MDP:
a strategy satisfying the multi-objective formula
on the finite-state region-graph MDP
provides a ``recipe'' for a corresponding region strategy on the semantic MDP.
For example, in the case of a single-objective threshold reachability formulae of the form 
$\probthresh{\thresh}{\lambda}(\Diamond \targetset)$ on finite MDPs, 
pure memoryless strategies are sufficient;
this allows us to conclude that pure memoryless region strategies suffice for PTA
with respect to single-objective threshold reachability formulae.
Finite-memory strategies are sufficient to establish 
that a finite-state MDP satisfies a multi-objective formula
of the following two classes:
\begin{inparaenum}[(1)]
\item
multiple-threshold formulae of the form
$\bigwedge_{i=1}^n \anobj_i(\anevent_i)$
for which
$\anobj_i \in 
\set{\probthresh{\thresh}{\lambda} \setsep \thresh \in \set{ >, \geq } \mbox{ and } \lambda \in [0,1] }$
and $\anevent_i$ is a reachability property for each $1 \leq i \leq n$
(from \cite{EKVY08});
\item
one-sure-parity one-threshold-reachability properties 
$\sure(\anevent_1) \wedge \probthresh{\thresh}{\lambda}(\anevent_2)$,
i.e., where $\anevent_1$ is a parity condition and $\anevent_2$
is a reachability property
(from \cite{BRR17}).
\end{inparaenum}
Hence finite-memory region strategies suffice to establish
such formulae on PTAs.
Conversely, 
we can conclude that memory and randomization are required 
by strategies of PTAs when considering these classes of formulae,
because they are required by finite-state MDP strategies on the same formula classes
(recall that PTAs conservatively extend MDPs,
i.e., a PTA with clock constraints that are always satisfied, regardless of the value of the clocks,
can be interpreted as an MDP).
Furthermore, 
infinite memory is required for determining whether finite-state MDPs
satisfy sure-threshold formulae of the form
$\sure(\anevent_1) \wedge \probthresh{\thresh}{\lambda}(\anevent_2)$,
where $\anevent_1$ and $\anevent_2$ are parity conditions  \cite{BRR17},
and hence this also applies to PTA strategies for such sure-threshold formulae.

In the following example, we present a PTA for which randomization \emph{over time delays}
is necessary for the satisfaction of a multi-objective formula.

\begin{example}\label{ex:rand_delays}
The PTA shown in \figref{fig:pta_nonzero},
which is a timed extension of the finite MDP of Figure~2 of \cite{BGR20},
satisfies the multi-objective formula 
$\probthresh{>}{0}(\Diamond \set{\aloc_2}) \wedge \probthresh{>}{0}(\Diamond \set{\aloc_3})$
by employing a strategy that assigns probability $\lambda$
to the action $(0,\apedge)$
and probability $1- \lambda$ to the action $(\adelay,\apedge)$,
for any probability $\lambda \in (0,1)$ and time delay $\adelay \in (0,1)$,
where $\apedge$ denotes the probabilistic edge with source location $\aloc_0$.
Strategies of the PTA that do not correspond to the above description do not satisfy the 
multi-objective formula.
\end{example}

\begin{figure}[!tb]
    \centering
    \begin{minipage}{.45\textwidth}
        \centering
\scriptsize
\begin{tikzpicture}[->,>=stealth',shorten >=1pt,auto, thin] 
  \tikzstyle{state}=[draw=black, text=black, shape=circle, inner sep=3pt, outer sep=0pt, circle, rounded corners] 
  \tikzstyle{final_state}=[draw=teal, text=teal, shape=circle, inner sep=3pt, outer sep=0pt, circle, rounded corners, fill= lg] 
  \tikzstyle{bad_state}=[draw=red, text=red, shape=circle, inner sep=3pt, outer sep=0pt, circle, rounded corners, fill= ly] 

	\node at (0,2) (for_initial) {};
	\node[state] at (0.5,2) (A) {$\ell_0$};
	\node[yshift=0.4cm,below of = A](inv_A){\colorbox{blue!7}{$\aclock < 1$}};
	\node[state] (B) at (2.2,2) {$\ell_1$};
	\node[xshift=-0.2cm,right of = B](inv_B){\colorbox{blue!7}{$\anotherclock \leq 0$}};
	\node[state] at (4,3.5) (C) {$\ell_2$};
	\node[yshift=0.4cm,below of = C](inv_C){\colorbox{blue!7}{$\aclock \leq 1$}};
	\node[state] at (4,0.5) (D) {$\ell_3$};
	\node[yshift=0.4cm,below of = D](inv_D){\colorbox{blue!7}{$\aclock \leq 1$}};

	\path[rounded corners]
	(for_initial)
		edge (A)
	(A)
		edge node[above] {\begin{tabular}{c}\colorbox{blue!7}{$\aclock < 1$}\\ \colorbox{blue!7}{$\set{\anotherclock}$}\end{tabular}} (B)
	(C)
		edge [loop above] node {\begin{tabular}{c}\colorbox{blue!7}{$\aclock = 1$}\\ \colorbox{blue!7}{$\set{\aclock}$}\end{tabular}} (C)
	(D)
		edge [loop above] node {\begin{tabular}{c}\colorbox{blue!7}{$\aclock = 1$}\\ \colorbox{blue!7}{$\set{\aclock}$}\end{tabular}} (D);

	\draw[->,rounded corners] (B.north) |- node[above,text=black,pos=.5]{\colorbox{blue!7}{$\aclock \! > \! 0 \wedge \anotherclock \! = \! 0$}} (C.west); 
	\draw[->,rounded corners] (B.south) |- node[below,text=black,pos=.5]{\colorbox{blue!7}{$\aclock \! = \! 0 \wedge \anotherclock \! = \! 0$}} (D.west); 

\end{tikzpicture}
        \caption{A PTA for which randomization over time delays is required (all edges correspond to probability {\colorbox{red!15}{$1$}}).}
        \label{fig:pta_nonzero}
    \end{minipage}%
	\hspace{0.5cm}
    \begin{minipage}{0.45\textwidth}
        \centering
\scriptsize
\begin{tikzpicture}[->,>=stealth',shorten >=1pt,auto, thin] 
  \tikzstyle{state}=[draw=black, text=black, shape=circle, inner sep=3pt, outer sep=0pt, circle, rounded corners] 
  \tikzstyle{final_state}=[draw=black, text=black, shape=circle, inner sep=3pt, outer sep=0pt, circle, rounded corners, fill= lg] 
  \tikzstyle{bad_state}=[draw=black, text=black, shape=circle, inner sep=3pt, outer sep=0pt, circle, rounded corners] 

	\node at (2.5,5) (for_initial){};
	\node at (2.5,4) [state](A) {$\aloc_0$};
	\node[yshift=0.2cm,below of = A](inv_A){\colorbox{blue!7}{$\aclock \leq 1$}};
	\node at (0,4) [fill=black] (nail_A) {};
	\node at (5,4) [state](B) {$\aloc_1$}; 
	\node[yshift=0.2cm,below of = B](inv_B){\colorbox{blue!7}{$\aclock \leq 1$}};

	\path[rounded corners]
	(for_initial)
		edge (A)
	(A)
		edge [bend left=60] node[above] {\begin{tabular}{c}{\colorbox{blue!7}{$\aclock \leq 1$}}\\{\colorbox{red!15}{$1$}}\end{tabular}} (B)
		edge node[above] {\colorbox{blue!7}{$\aclock \leq 1$}} (nail_A)
	(nail_A)
		edge [bend left=60] node[above,text=black,pos=.5] {\colorbox{red!15}{$\frac{1}{2}$}}  (A)
		edge [bend right=60] node[below,text=black,pos=.5] {\colorbox{red!15}{$\frac{1}{2}$}\colorbox{blue!7}{$\set{\aclock}$}} (A)
	(B)
		edge [bend left=60] node[below] {\begin{tabular}{c}{\colorbox{blue!7}{$\aclock \leq 1$}}\\{\colorbox{red!15}{$1$}} \colorbox{blue!7}{$\set{\aclock}$}\end{tabular}} (A);


%
%
%
%
%
%


\end{tikzpicture}
        \caption{A PTA adapted from Example~1 of \cite{BRR17}, for which infinite memory is required to choose between probabilistic edges.}
        \label{fig:pta_brr17}
    \end{minipage}
\end{figure}

\subsubsection{Delay-restricted strategies.}
Although Example~\ref{ex:rand_delays}
shows that randomization over time delays can be necessary for satisfying multi-objective formulae,
we now show that randomization between time delays 
leading to the \emph{same clock equivalence class}
is not necessary:
for any strategy randomizing between delays leading to the same clock equivalence class,
we can find an equivalent strategy
that assigns positive probability to at most one delay in that class.
This result is key for the proof of \propref{prop:mof_rg}.

A strategy $\astrat \in \strats{\semMDP{\PTA}}$
is \emph{delay-restricted} if,
for all $\afinpath \in \mfinpaths{\semMDP{\PTA}}$,
and each distinct $(\adelay,\apedge),(\adelay',\apedge') \in \support(\astrat(\afinpath))$,
either $\apedge \neq \apedge'$ or 
$\cecof{\valof{\last(\afinpath)} + \adelay} \neq \cecof{\valof{\last(\afinpath)} + \adelay'}$.
Let $\drstrats{\semMDP{\PTA}} \subseteq \strats{\semMDP{\PTA}}$
be the set of delay-restricted strategies of $\semMDP{\PTA}$.

For a strategy $\astrat \in \strats{\semMDP{\PTA}}$ 
and finite region path $\afinpathreg \in \mfinpaths{\regMDP{\PTA}}$,
let $\dfinpathsfromregpath{\astrat}{\afinpathreg} 
= 
\set{ \afinpath \in \mfinpaths{\astrat} \setsep \afinpath \in \afinpathreg}$.
We recall the following definitions.
First, given a strategy $\astrat \in \strats{\semMDP{\PTA}}$,
we define the induced MC 
$\adtmcparaone{\astrat} = (\mfinpaths{\astrat},\minit,\dtransof{\astrat})$.
Second, for any finite path $\afinpath \in \dfinpaths{\adtmcparaone{\astrat}}$,
we use $\probfin{\adtmcparaone{\astrat}}(\afinpath)$,
to denote the probability of $\afinpath$ in $\adtmcparaone{\astrat}$.
Third, we recall the 1-to-1 relationship between paths generated by $\astrat$
and paths of $\adtmcparaone{\astrat}$.
For any finite path $\afinpath \in \mfinpaths{\astrat}$
we write $\dtmcpathof{\astrat}{\afinpath}$ to be the unique finite path 
of $\dfinpaths{\adtmcparaone{\astrat}}$ 
according to this relationship.
The above allows us to define the function $\probfin{\astrat}: \mfinpaths{\astrat} \ra [0,1]$
mapping paths generated by $\astrat$ to probabilities:
for each $\afinpath \in \mfinpaths{\astrat}$,
let $\probfin{\astrat}(\afinpath) = \probfin{\adtmcparaone{\astrat}}(\dtmcpathof{\astrat}{\afinpath})$.
Let $\probfin{\astrat}(\dfinpathsfromregpath{\astrat}{\afinpathreg})
=
\sum_{\afinpath \in \dfinpathsfromregpath{\astrat}{\afinpathreg}} 
\probfin{\astrat}(\afinpath)$.

The following lemma specifies that,
for any strategy, we can find an delay-restricted strategy
that assigns the same probabilities to paths corresponding 
to the same finite region paths.
\begin{lemma}\label{lem:dr_finpaths}
Let $\astrat \in \strats{\semMDP{\PTA}}$ be a strategy of $\semMDP{\PTA}$.
Then there exists a delay-restricted strategy $\adrstrat \in \drstrats{\semMDP{\PTA}}$
such that, 
for each finite region path $\afinpathreg$ of $\regMDP{\PTA}$,
we have $\probfin{\astrat}(\dfinpathsfromregpath{\astrat}{\afinpathreg})
=
\probfin{\adrstrat}(\dfinpathsfromregpath{\adrstrat}{\afinpathreg})$.
\end{lemma}
The proof of \lemref{lem:dr_finpaths}
involves constructing the delay-restricted strategy $\adrstrat$
from the standard strategy $\astrat$
by transforming a choice over multiple time delays leading to the same clock equivalence class
into a single time delay to the same clock equivalence class.
From \lemref{lem:dr_finpaths},
we can use results on probability measures 
(similar to those used in proof of Lemma~10.66 in \cite{BK08}
in the context of probabilistic bisimulation),
to conclude that the probabilities assigned by a standard strategy 
$\astrat$
and its associated delay-restricted strategy $\adrstrat$
to events are equal.
Furthermore, given that $\astrat$ and $\adrstrat$ exhibit the same set of traces,
they satisfy the same sure objectives.
Overall we obtain the following result.

\begin{lemma}\label{lem:dr_mof}
Let $\mof$ be a multi-objective formula with respect to the set of locations of $\PTA$.
If there exists a strategy $\astrat \in \strats{\semMDP{\PTA}}$
such that $\objsat{(\semMDP{\PTA},\mdplabof{\semMDP{\PTA}})}{\astrat}{\mof}$,
then there exists a delay-restricted strategy $\adrstrat \in \drstrats{\semMDP{\PTA}}$
such that $\objsat{(\semMDP{\PTA},\mdplabof{\semMDP{\PTA}})}{\adrstrat}{\mof}$. 
\end{lemma}

\subsubsection{Correctness of \propref{prop:mof_rg}.}
We now use \lemref{lem:dr_mof}, 
together with the two lemmata below,
to establish \propref{prop:mof_rg}.
First, we show that $\objsatpta{\PTA}{\mof}$ 
implies
$\objsatmdp{(\regMDP{\PTA},\mdplabof{\regMDP{\PTA}})}{\mof}$.
Recalling that $\objsatpta{\PTA}{\mof}$ indicates that 
there exists a strategy $\astrat \in \strats{\semMDP{\PTA}}$
such that $\objsat{(\semMDP{\PTA},\mdplabof{\semMDP{\PTA}})}{\astrat}{\mof}$,
we use \lemref{lem:dr_mof}
to identify a delay-restricted strategy $\adrstrat \in \drstrats{\semMDP{\PTA}}$
such that $\objsat{(\semMDP{\PTA},\mdplabof{\semMDP{\PTA}})}{\adrstrat}{\mof}$.
Then we use \lemref{lem:dr_to_reg} (stated below) and \lemref{lem:equiv_mdp_mof}
to conclude that there exists 
a strategy $\aregstratof{\adrstrat} \in \strats{\regMDP{\PTA}}$ 
such that $\objsat{(\regMDP{\PTA},\mdplabof{\regMDP{\PTA}})}{\aregstratof{\adrstrat}}{\mof}$,
and hence $\objsatmdp{(\regMDP{\PTA},\mdplabof{\regMDP{\PTA}})}{\mof}$.
\lemref{lem:dr_to_reg} 
uses \lemref{lem:reset_cor} to ensure that the probability distributions over next states
from each related 
(by the bijection witnessing isomorphism) 
pair of states of $\adtmcparaone{\adrstrat}$ and $\adtmcparaone{\aregstratof{\adrstrat}}$ 
have the required relationship necessary for isomorphism.

\begin{lemma}\label{lem:dr_to_reg}
Let $\adrstrat \in \drstrats{\semMDP{\PTA}}$ be a delay-restricted strategy of $\semMDP{\PTA}$.
Then there exists a strategy $\aregstratof{\adrstrat} \in \strats{\regMDP{\PTA}}$ of $\regMDP{\PTA}$ 
such that the labeled MCs 
$(\adtmcparaone{\adrstrat},\dtmclabfrom{\mdplabof{\semMDP{\PTA}}})$ 
and 
$(\adtmcparaone{\aregstratof{\adrstrat}},\dtmclabfrom{\mdplabof{\regMDP{\PTA}}})$
are isomorphic.
\end{lemma}

The reverse direction of \propref{prop:mof_rg} requires showing that 
$\objsatmdp{(\regMDP{\PTA},\mdplabof{\regMDP{\PTA}})}{\mof}$
implies $\objsatpta{\PTA}{\mof}$.
Recall that 
$\objsatmdp{(\regMDP{\PTA},\mdplabof{\regMDP{\PTA}})}{\mof}$
indicates the existence of a strategy $\astrat \in \strats{\regMDP{\PTA}}$ of $\regMDP{\PTA}$
such that $\objsat{(\regMDP{\PTA},\mdplabof{\regMDP{\PTA}})}{\astrat}{\mof}$.
Then the combination of the following lemma and \lemref{lem:equiv_mdp_mof}
implies the existence of $\astrat' \in \strats{\semMDP{\PTA}}$
such that $\objsat{(\semMDP{\PTA},\mdplabof{\semMDP{\PTA}})}{\astrat'}{\mof}$,
from which we conclude that $\objsatpta{\PTA}{\mof}$

\begin{lemma}\label{lem:rg_to_sem}
Let $\astrat \in \strats{\regMDP{\PTA}}$ be a strategy of $\regMDP{\PTA}$.
Then there exists a strategy $\astrat' \in \strats{\semMDP{\PTA}}$ of $\semMDP{\PTA}$ 
such that the labeled MCs 
$(\adtmcparaone{\astrat},\dtmclabfrom{\mdplabof{\regMDP{\PTA}}})$
and
$(\adtmcparaone{\astrat'},\dtmclabfrom{\mdplabof{\semMDP{\PTA}}})$  
are isomorphic.
\end{lemma}

\subsubsection{Complexity of verifying multi-objective formulae on PTAs.}
We now identify upper bounds on the complexity 
of the problem of determining whether a multi-objective formula
is satisfied by a PTA,
for certain subclasses of formulae.
These results follow from the combination of the following:
\propref{prop:mof_rg},
the size of the region-graph MDP being exponential in the size of the encoding of the PTA,
and previous work on subclasses of multi-objective formulae.
For context, we recall that the classical result of \cite{KNSS02}
provides an approach for single-objective threshold reachability formulae of the form 
$\probthresh{\thresh}{\lambda}(\Diamond \targetset)$
based on standard probabilistic reachability analysis of the region-graph MDP.
Such analysis can be done in $\ptime$ for finite-state MDPs \cite{deAlfPhD,CY98},
and hence determining whether a single-objective threshold reachability formulae 
is satisfied by a PTA can be done in $\exptime$.
Consider multi-objective threshold formulae of the form
$\bigwedge_{i=1}^n \anobj_i(\anevent_i)$
for which
$\anobj_i \in 
\set{\probthresh{\thresh}{\lambda} \setsep \thresh \in \set{ >, \geq } \mbox{ and } \lambda \in [0,1] }$
and $\anevent_i$ is a sink-reachability property 
for each $1 \leq i \leq n$.
Such formulae can be verified on finite-state MDPs in polynomial time
\cite{EKVY08}.
Hence establishing whether multi-objective formulae of this form
are satisfied for a PTA can be done in $\exptime$.
Single-sure-single-threshold formulae of the form
$\sure(\anevent_1) \wedge \probthresh{\thresh}{\lambda}(\anevent_2)$,
where $\anevent_1$ is a parity condition 
and $\anevent_2$ is a reachability property or a  parity condition,
can be verified on finite-state MDPs in $\npconp$ \cite{BRR17},
and hence in $\ncoexptime$ on PTAs.
Finally, consider 
single-sure-multiple-threshold formulae of the form
$\sure(\anevent) \wedge \bigwedge_{i=1}^n \anobj_i(\anevent_i)$,
where $\anevent$ is a parity condition 
and, for each $1 \leq i \leq n$, 
$\anobj_i \in 
\set{\probthresh{\thresh}{\lambda} \setsep \thresh \in \set{ >, \geq } \mbox{ and } \lambda \in [0,1] }$
and $\anevent_i$ is sink reachability property.
Such formulae
can be verified on finite-state MDPs in $\exptime$ \cite{BKW24}, 
and therefore in $\twoexptime$ on PTAs.
All of the above problems are $\exptime$-hard because they subsume
the single-objective threshold reachability case \cite{JLS08}.

\section{Time Divergence and Multi-objective Formulae}\label{sec:divergence}
We now focus on the issue of time divergence in PTA,
and show how conjuncts of multi-objective formulae can restrict
formula evaluation to divergent strategies.

A path $\apath = \astate_0 (\adelay_0,\apedge_0) \astate_1 (\adelay_1,\apedge_1) \ldots$ 
of $\semMDP{\PTA}$
is \emph{divergent}
if, for each $n \in \Nset$,
there exists $i \in \Nset$ such that 
$\sum_{j=0}^{i} \adelay_j > n$.
Let $\divpaths \subseteq \mpaths{\semMDP{\PTA}}$ be 
the set of divergent paths of $\semMDP{\PTA}$.
A strategy $\astrat \in \strats{\semMDP{\PTA}}$
is \emph{almost-sure divergent} if $\dprob{\astrat}(\divpaths) = 1$
and is \emph{sure divergent} if $\mpaths{\astrat} \subseteq \divpaths$.
Let $\asstrats{\semMDP{\PTA}}$ be the set of almost-sure divergent strategies,
and $\surestrats{\semMDP{\PTA}}$ be the set of sure divergent strategies.
Furthermore, we write $\asobjsatpta{\PTA}{\mof}$ 
(respectively, $\sureobjsatpta{\PTA}{\mof}$) if and only if 
there exists an almost-sure strategy $\astrat \in \asstrats{\semMDP{\PTA}}$
(respectively, sure strategy $\astrat \in \surestrats{\semMDP{\PTA}}$) 
such that $\objsat{(\semMDP{\PTA},\mdplabof{\semMDP{\PTA}})}{\astrat}{\mof}$.

We now show how a PTA can be expanded to represent information
concerning whether total elapsed time has reached a natural number,
using ideas from \cite{dAFHMS03,AH97} and the precedent of \cite{Spr09}.
Given a PTA $\PTA = (\locs, \locinit, \clocks, \inv, \pedges)$,
its \emph{tick-expanded PTA} is defined 
in the following way.
First, for any location $\aloc \in \locs$,
we write $\anexloc$ for its corresponding \emph{tick location}.
In the following construction, no time elapses in tick locations,
and a tick location is visited at exactly $1$ time unit after the
a tick location was last visited (or after the start of the execution of the PTA).
We write $\ticklocs = \set{ \anexloc \setsep \aloc \in \locs}$
for the set of tick locations.
Then we define the tick-expanded PTA as
$\exPTA = (\locs \cup \ticklocs, \locinit, \clocks \cup \set{\exclock}, \exinv, \expedges)$,
where 
$\exinv(\aloc) = \exinv(\anexloc) = \inv(\aloc) \wedge (\exclock \leq 1)$ for each $\aloc \in \locs$,
and $\expedges = \pedges 
\cup 
\set{(\aloc,(\exclock=1),\pd_{\emptyset,\anexloc}), (\anexloc,(\exclock=1),\pd_{\set{\exclock},\aloc})
\setsep \aloc \in \locs}$,
where $\pd_{\emptyset,\anexloc}$ and $\pd_{\set{\exclock},\aloc}$
denote the distributions such that $\pd_{\emptyset,\anexloc}(\emptyset,\anexloc) = 1$
and $\pd_{\set{\exclock},\aloc}(\set{\exclock},\aloc) = 1$, respectively.
In the case of a parity condition,
the priority function $\prfn: \locs \ra \set{1,2,\ldots,d}$ on $\PTA$ 
is represented by the \emph{expanded priority function} $\exprfn: \locs \cup \ticklocs \ra \set{1,2,\ldots,d}$ 
on $\exPTA$,
where $\exprfn(\aloc) = \exprfn(\anexloc) = \prfn(\aloc)$ for each $\aloc \in \locs$.

Note that the tick-expanded PTA $\exPTA$ does not change the behavior of $\PTA$
with regard to the satisfaction of reachability properties 
or (with the definition of the expanded parity function above) parity conditions:
the tick locations, their incoming and outgoing edges and clock constraints,
act merely as an observation mechanism.
For any multi-objective formula 
$\mof = \bigwedge_{i=1}^n \anobj_i(\anevent_i)$ 
with respect to the locations of $\PTA$,
we define $\exmof = \bigwedge_{i=1}^n \anobj_i(\exanevent_i)$ 
to be the multi-objective formula with respect to the locations of $\exPTA$,
where, for $1 \leq i \leq n$,
if $\anevent_i$ is $\parity{\prfn_i}$ then $\exanevent_i$ is $\parity{\exprfn_i}$,
and if  $\anevent_i$ is $\Diamond \targetset_i$ then $\exanevent_i$ is also $\Diamond \targetset_i$
(i.e., the parity functions featured in $\exmof$ are the expanded parity functions 
derived from those of $\mof$, as described above,
and reachability properties featured in $\exmof$ remain the same as in $\mof$).
From the definitions of $\exPTA$ and $\exmof$,
we have that $\objsatpta{\PTA}{\mof}$ if and only if $\objsatpta{\exPTA}{\exmof}$.
Next, observe that a semantic MDP path of $\exPTA$ is divergent if and only if 
the path visits a tick location infinitely often.
We use $\tickio$ to represent the set of traces visiting tick location infinitely often,
both in $\exPTA$ and in $\regMDP{\exPTA}$.
Note that $\tickio$ can be represented by the priority function 
$\tickprfn: \locs \cup \ticklocs \ra \set{1,2}$
where $\tickprfn(\aloc) = 1$ and $\tickprfn(\anexloc) = 2$ for each $\aloc \in \locs$
(note that $\tickprfn$ expresses a B\"{u}chi property).
From the above reasoning, and from \propref{prop:mof_rg}, 
we conclude that 
both almost-sure and sure divergence can be represented as an additional conjunct 
when analyzing the region-graph MDP.

\begin{proposition}\label{prop:td}
Let $\PTA$ be a PTA  
and $\mof$ be a multi-objective formula with respect to the set of locations of $\PTA$.
Then $\asobjsatpta{\PTA}{\mof}$ 
if and only if 
$\objsatmdp{(\regMDP{\exPTA},\mdplabof{\regMDP{\exPTA}})}
{\mof \wedge \probthresh{\geq}{1}(\tickio)}$,
and 
$\sureobjsatpta{\PTA}{\mof}$ 
if and only if 
$\objsatmdp{(\regMDP{\exPTA},\mdplabof{\regMDP{\exPTA}})}{\mof \wedge \sure(\tickio)}$.
\end{proposition}
The case of a single probabilistic threshold parity objective 
while restricting to sure divergence strategies is a new result,
based on the use of the algorithms of \cite{BRR17},
because previous results on sure divergence considered only threshold properties
for reachability and safety objectives \cite{Spr09}.
For multi-objective threshold sink-reachability properties and almost-sure divergence
we obtain an algorithm in \exptime~on inspection of the construction 
of the proof of \cite[Theorem~5.1(1)]{EKVY08}.

\begin{example}
Consider the PTA of \figref{fig:pta_brr17},
and the single-objective formula $\probthresh{\geq}{1}(\anevent_\prfn)$
where $\anevent_\prfn$ is an event based on the priority function $\prfn$
defined by $\prfn(\aloc_0) = 0$ and  $\prfn(\aloc_1) = 1$;
that is, the formula expresses the coB\"{u}chi property 
that $\aloc_1$ should be visited only finitely often.
This example is a PTA version of the MDP of Example~1 of \cite{BRR17},
which requires infinite memory to satisfy a sure-B\"{u}chi threshold-coB\"{u}chi formula.
In our setting, we use sure-divergence instead of a sure B\"{u}chi objective.
Note that the single-objective formula $\probthresh{\geq}{1}(\anevent_\prfn)$ 
can be satisfied by sure-divergent region strategies,
but these strategies require infinite memory to make the choice between the leftwards
and rightwards edges from $\aloc_0$.
Such a strategy operates in rounds:
in round $i$, the strategy lets $1$ time unit elapse,
then chooses to visit the black square repeatedly, 
aiming to reset the clock $\aclock$, with a number of attempts
such that the probability of resetting $\aclock$ exceeds $1-\frac{1}{2^i}$.
The strategy chooses to visit $\aloc_1$ only when the current round's 
number of attempts is exceeded without $\aclock$ having been reset.
After resetting clock $\aclock$, the strategy enters round $i+1$,
and hence the number of attempts it will make to reset $\aclock$
via the black square is increased.
This strategy guarantees that time diverges on all of its paths,
and that visiting $\aloc_1$ infinitely often has probability $0$.

Note that the infinite memory required for this example refers to choice between
(probabilistic) edges.
This contrasts to Example~\ref{ex:ptadiv},
in which $\probthresh{\geq}{1}(\Diamond \set{\aloc_2})$
is satisfied (by general or almost-sure strategies)
by choosing repeatedly the rightwards probabilistic edge from $\aloc_0$ until $\aloc_1$ is reached,
but by choosing progressively shorter time delays in $\aloc_0$ so that 
the value of $\aclock$ does not exceed $1$;
hence infinite memory is required in Example~\ref{ex:ptadiv} to choose appropriate time delays.
\end{example}

\section{Conclusion}
We have seen how multi-objective properties for PTA can be verified,
by relying on the region-graph MDP construction 
and existing methods for multi-objective properties.
For future work, 
we conjecture that sure-divergence single-threshold formulae
can be verified in $\exptime$,
rather than in $\ncoexptime$,
based on the fact that a fixed number of priorities can be used to reason about divergence
on the enlarged PTA,
and considering the results of \cite{DG26}
in the qualitative setting.
We also conjecture that allowing strategies to randomize over uncountable sets of time delays,
rather than finite sets as considered in this paper,
gives no advantage in the PTA and multi-objective setting.
Recalling that \cite{EKVY08} provide algorithms 
also for multi-objective parity formulae of finite-state MDPs,
future work could consider the complexity of checking such formulae on PTA.
Extensions to rewards \cite{FKNPQ11},
deterministic timed automata objectives (generalizing \cite{FLL18}) 
more general Boolean combinations (as in \cite{BGR20} for qualitative properties),
and approximation of the Pareto frontier for multiple objectives
can also be considered in the PTA setting.
Multiple-objective threshold properties 
form the basis of the probabilistic strategy logic of \cite{JKPPS26},
and the results of this paper could potentially have relevance for the development of 
algorithms and complexity results
for that logic.

\begin{credits}
\subsubsection{\ackname} This work has received funding from the European Union’s
Horizon research and innovation program Chips JU under
Grant Agreement No. 101139769, DistriMuSe project
(HORIZON-KDT-JU-2023-2-RIA). The JU receives support
from the European Union’s Horizon research and innovation
programme and the nations involved in the mentioned
projects. The work reflects only the author’s views; the European
Commission is not responsible for any use that may
be made of the information it contains. 
\end{credits}

 \bibliographystyle{splncs04}
 \bibliography{sure_pta}

@preamble{ {\providecommand{\JP}{J.-P} } }

@preamble{ {\providecommand{\JF}{J.-F} } }

@String{tcs =    "Theoretical Computer Science"}

@String{fmsd =    "Formal Methods in System Design"}

@article{AD94,
        author="R. Alur and D. L. Dill",
        title="A theory of timed automata",
        journal=tcs,
        year={1994},
        volume={126},
        number="2",
        pages="183-235",
  doi          = {10.1016/0304-3975(94)90010-8},
}

@article{AH97,
  author       = {R. Alur and T. Henzinger},
  title        = {Real-Time System = Discrete System + Clock Variables},
  journal      = {Int. J. Softw. Tools Technol. Transf.},
  volume       = {1},
  number       = {1-2},
  pages        = {86--109},
  year         = {1997},
  doi          = {10.1007/S100090050007}
}

@inproceedings{BGR20,
  author       = {R. Berthon and S. Guha and {\JF} Raskin},
  title        = {Mixing Probabilistic and non-Probabilistic Objectives in {M}arkov Decision
                  Processes},
  booktitle    = {Proc. LICS 2020},
  pages        = {195--208},
  publisher    = {{ACM}},
  year         = {2020},
  doi          = {10.1145/3373718.3394805}
}

@inproceedings{BKW24,
  author       = {R. Berthon and {\JP} Katoen and T. Winkler},
  title        = {Markov Decision Processes with Sure Parity and Multiple Reachability
                  Objectives},
  booktitle    = {Proc. RP 2024},
  series       = {LNCS},
  volume       = {15050},
  pages        = {203--220},
  publisher    = {Springer},
  year         = {2024},
  doi          = {10.1007/978-3-031-72621-7\_14}
}

@inproceedings{BRR17,
  author       = {R. Berthon and M. Randour and {\JF} Raskin},
  title        = {Threshold Constraints with Guarantees for Parity Objectives in {M}arkov
                  Decision Processes},
  booktitle    = {Proc. ICALP 2017},
  series       = {LIPIcs},
  volume       = {80},
  pages        = {121:1--121:15},
  publisher    = {Schloss Dagstuhl - Leibniz-Zentrum f{\"{u}}r Informatik},
  year         = {2017},
  doi          = {10.4230/LIPICS.ICALP.2017.121}
}

@book{BK08,
  author    = {C. Baier and
               {\JP} Katoen},
  title     = {Principles of model checking},
  publisher = {{MIT} Press},
  year      = {2008}
}

@inproceedings{CHP08,
  author       = {K. Chatterjee and T. Henzinger and V. Prabhu},
  title        = {Trading Infinite Memory for Uniform Randomness in Timed Games},
  booktitle    = {Proc. HSCC 2008},
  series       = {LNCS},
  volume       = {4981},
  pages        = {87--100},
  publisher    = {Springer},
  year         = {2008},
  doi          = {10.1007/978-3-540-78929-1\_7}
}

@inproceedings{CMH06,
  author       = {K. Chatterjee and R. Majumdar and T. Henzinger},
  title        = {Markov Decision Processes with Multiple Objectives},
  booktitle    = {Proc. STACS 2006},
  series       = {LNCS},
  volume       = {3884},
  pages        = {325--336},
  publisher    = {Springer},
  year         = {2006},
  doi          = {10.1007/11672142\_26}
}

@InProceedings{CKSW13,
author={T. Chen and M. Kwiatkowska and A. Simaitis and C. Wiltsche},
title={Synthesis for Multi-Objective Stochastic Games: An Application to Autonomous Urban Driving},
booktitle={Proc. QEST 2013)},
year={2013},
pages={322--337},
publisher={Springer},
series={LNCS},
volume={8054}
}

@article{CY98,
  author       = {C Courcoubetis and M. Yannakakis},
  title        = {Markov decision processes and regular events},
  journal      = {{IEEE} Trans. Autom. Control.},
  volume       = {43},
  number       = {10},
  pages        = {1399--1418},
  year         = {1998},
  doi          = {10.1109/9.720497}
}

@phdthesis{deAlfPhD,
  title = {Formal Verification of Probabilistic Systems},
  author = {L. de Alfaro},
  school = {Stanford University},
  year = {1997}
  }

@inproceedings{dAFHMS03,
  author       = {L. de Alfaro and
                  M. Faella and
                  T. Henzinger and
                  R. Majumdar and
                  M. Stoelinga},
  title        = {The Element of Surprise in Timed Games},
  booktitle    = {Proc. CONCUR 2003},
  series       = {LNCS},
  volume       = {2761},
  pages        = {142--156},
  publisher    = {Springer},
  year         = {2003},
  doi          = {10.1007/978-3-540-45187-7\_9}
}

@inproceedings{DG26,
  author       = {L. Doyen and S. Guha},
  title        = {Algorithm and Strategy Construction for Sure-Almost-Sure Stochastic
                  Parity Games},
  booktitle    = {Proc. STACS 2026},
  series       = {LIPIcs},
  volume       = {364},
  pages        = {34:1--34:20},
  publisher    = {Schloss Dagstuhl - Leibniz-Zentrum f{\"{u}}r Informatik},
  year         = {2026},
  doi          = {10.4230/LIPICS.STACS.2026.34}
}

@article{EKVY08,
  author       = {K. Etessami and M. Kwiatkowska and M. Vardi and M. Yannakakis},
  title        = {Multi-Objective Model Checking of {M}arkov Decision Processes},
  journal      = {Logical Methods in Comput. Scence.},
  volume       = {4},
  number       = {4},
  year         = {2008},
  doi          = {10.2168/LMCS-4(4:8)2008}
}

@InProceedings{FKP12,
author={V. Forejt and M. Kwiatkowska and D. Parker},
title={Pareto Curves for Probabilistic Model Checking},
booktitle={Proc. ATVA 2012},
year={2012},
pages={317--332},
publisher={Springer},
series={LNCS},
volume={7561},
doi          = {10.1007/978-3-642-33386-6\_25}
}

@inproceedings{FKNPQ11,
  author       = {V. Forejt and M. Kwiatkowska and G. Norman and D. Parker and H. Qu},
  title        = {Quantitative Multi-objective Verification for Probabilistic Systems},
  booktitle    = {Proc. TACAS 2011},
  series       = {LNCS},
  volume       = {6605},
  pages        = {112--127},
  publisher    = {Springer},
  year         = {2011},
  doi          = {10.1007/978-3-642-19835-9\_11}
}

@inproceedings{FLL18,
  author       = {Hongfei Fu and Yi Li and Jianlin Li},
  title        = {Verifying Probabilistic Timed Automata Against Omega-Regular Dense-Time
                  Properties},
  booktitle    = {Proc. QEST 2018},
  series       = {LNCS},
  volume       = {11024},
  pages        = {122--139},
  publisher    = {Springer},
  year         = {2018},
  doi          = {10.1007/978-3-319-99154-2\_8}
}

@inproceedings{HQvW26,
  author       = {A. Hartmanns and T. Quatmann and M. van Wijk},
  title        = {Tools and Algorithms for Sound Multi-Objective Probabilistic Model
                  Checking - (Long Tool Paper)},
  booktitle    = {Proc. FM 2026},
  series       = {LNCS},
  volume       = {16556},
  pages        = {189--210},
  publisher    = {Springer},
  year         = {2026},
  doi          = {10.1007/978-3-032-26204-2\_10}
}

@mastersthesis{GJ95,
	author =	{H. Gregersen and H. E. Jensen},
	school =	{Department of Mathematics and Computer Science, Aalborg University},
	type =	{Master's Thesis},
	title =	{Formal Design of Reliable Real Time Systems},
	year =	{1995},
}

@inproceedings{HH14,
  author       = {A. Hartmanns and H. Hermanns},
  title        = {The {{M}odest} Toolset: An Integrated Environment for Quantitative Modelling
                  and Verification},
  booktitle    = {Proc. TACAS 2014},
  series       = {LNCS},
  volume       = {8413},
  pages        = {593--598},
  publisher    = {Springer},
  year         = {2014},
  doi          = {10.1007/978-3-642-54862-8\_51}
}

@article{HHH14,
  author       = {E.M. Hahn and A. Hartmanns and H. Hermanns},
  title        = {Reachability and Reward Checking for Stochastic Timed Automata},
  note   = {{P}roc. AvoCS 2014},
  journal      = {Electron. Commun. Eur. Assoc. Softw. Sci. Technol.},
  volume       = {70},
  year         = {2014},
  doi          = {10.14279/TUJ.ECEASST.70.968}
}

@inproceedings{JKPPS25,
  author       = {W. Jamroga and M. Kwiatkowska and W. Penczek and L. Petrucci and T. Sidoruk},
  title        = {Probabilistic Timed {ATL}},
  booktitle    = {Proc. AAMAS 2025},
  pages        = {1051--1059},
  publisher    = {{IFAAMAS}/{ACM}},
  year         = {2025},
  doi          = {10.5555/3709347.3743625}
}

@inproceedings{JKPPS26,
  author       = {W. Jamroga and M. Kwiatkowska and W. Penczek and L. Petrucci and T. Sidoruk},
  title        = {Towards Probabilistic Strategic Timed {CTL}},
  booktitle    = {Proc. AAMAS 2026},
  pages        = {3468--3470},
  publisher    = {{IFAAMAS}/{ACM}},
  year         = {2026},
  doi          = {10.65109/OEXM4130}
}

@Article{JLS08,
  author =       {M. Jurdzi\'{n}ski and F. Laroussinie and J. Sproston},
  title =        {Model Checking Probabilistic Timed Automata with One or Two Clocks},
  journal =      {Logical Methods in Computer Science},
  year =         {2008},
  volume =       {4},
  number =       {3},
  doi          = {10.2168/LMCS-4(3:12)2008}
}

@inproceedings{KNP11,
author={M. Kwiatkowska and G. Norman and D. Parker},
title={{PRISM} 4.0: Verification of Probabilistic Real-time Systems},
booktitle={Proc. CAV 2011},
year={2011},
pages={585--591},
publisher={Springer},
series={LNCS},
volume={6806},
doi = {10.1007/978-3-642-22110-1_47}
}

@article{KNPS06,
author={M. Kwiatkowska and G. Norman and D. Parker and J. Sproston},
title={Performance Analysis of Probabilistic Timed Automata using Digital Clocks},
journal=fmsd,
year={2006},
volume={29},
pages={33-78},
doi       = {10.1007/s10703-006-0005-2}
}

@Article{KNSS02,
  author =       {M. Kwiatkowska and G. Norman and R. Segala and J. Sproston},
  title =        {Automatic verification of real-time systems with discrete probability distributions},
  journal =      tcs,
  year =         {2002},
  volume =       {286},
  pages =        {101-150},
  doi          = {10.1016/S0304-3975(01)00046-9}
}

@InProceedings{LPH17,
author={B. Lacerda and D. Parker and N. Hawes},
title={Multi-objective Policy Generation for Mobile Robots under Probabilistic Time-Bounded Guarantees},
booktitle={Proc. ICAPS 2017},
year={2017},
pages={504--512},
publisher={AAAI Press}
}

@Article{NPS13,
author={G. Norman and D. Parker and J. Sproston},
title={Model Checking for Probabilistic Timed Automata},
journal={Formal Methods in System Design},
year={2013},
volume={43},
number={2},
pages={164--190},
publisher={Springer},
  doi          = {10.1007/S10703-012-0177-X}
}

@inproceedings{Spr09,
  author       = {J. Sproston},
  title        = {Strict Divergence for Probabilistic Timed Automata},
  booktitle    = {Proc. CONCUR 2009},
  series       = {LNCS},
  volume       = {5710},
  pages        = {620--636},
  publisher    = {Springer},
  year         = {2009},
  doi          = {10.1007/978-3-642-04081-8\_41}
}

@inproceedings{Spr11,
  author       = {J. Sproston},
  title        = {Discrete-Time Verification and Control for Probabilistic Rectangular
                  Hybrid Automata},
  booktitle    = {Proc. QEST 2011},
  pages        = {79--88},
  publisher    = {{IEEE} Computer Society},
  year         = {2011},
  doi          = {10.1109/QEST.2011.18}
}

@article{Spr21FI,
  author    = {J. Sproston},
  title     = {Probabilistic Timed Automata with Clock-Dependent Probabilities},
journal={Fundamenta Informaticae},
year={2021},
volume={178},
number={1--2},
pages={101--138},
publisher={IOS Press},
  doi          = {10.3233/FI-2021-2000}
}

\section{Appendix}
\subsection{Proof of \lemref{lem:equiv_mdp_mof}}

\begin{proof}
Given that $\objsatmdp{(\anmdp_1,\mdplab_1)}{\mof}$,
there exists exists $\astrat_1 \in \strats{\anmdp_1}$
such that $\objsat{(\anmdp_1,\mdplab_1)}{\astrat_1}{\mof}$.
The statement of the lemma specifies that
there exists $\astrat_2 \in \strats{\anmdp_2}$
such that $(\adtmcparaone{\astrat_1},\dtmclabfrom{\mdplab_1})$
and $(\adtmcparaone{\astrat_2},\dtmclabfrom{\mdplab_2})$ are isomorphic.
From \lemref{lem:isopath} and \lemref{lem:isoprob},
we have that, for each $1 \leq i \leq n$,
if $\objsat{(\anmdp_1,\mdplab_1)}{\astrat_1}{\anobj_i(\anevent_i)}$
then $\objsat{(\anmdp_2,\mdplab_2)}{\astrat_2}{\anobj_i(\anevent_i)}$.
Hence 
$\objsat{(\anmdp_2,\mdplab_2)}{\astrat_2}{\mof}$.
Thus we have identified a strategy 
from which we obtain $\astrat_2 \in \strats{\anmdp_2}$ 
such that $\objsat{(\anmdp_2,\mdplab_2)}{\astrat_2}{\mof}$,
from which we conclude that 
$\objsatmdp{(\anmdp_2,\mdplab_2)}{\mof}$.
\qed
\end{proof}

\subsection{Technical lemma required for \lemref{lem:dr_finpaths}}

The following lemma specifies that, 
for a delay-restricted strategy $\astrat \in \drstrats{\semMDP{\PTA}}$,
and for any finite region path $\afinpathreg$,
there exists at most one path $\afinpath \in \mfinpaths{\astrat}$
such that $\afinpath \inregpath \afinpathreg$.
Instead, for a general (non-delay-restricted) strategy, this property generally does not hold.
In the following, 
for any probabilistic edge $\apedge \in \pedges$ 
where $\apedge = (\aloc,\g,\pd)$,
then $\grdof{\apedge} = \g$ and $\distof{\apedge} = \pd$.

\begin{lemma}\label{lem:drstrat_one_path}
Given a delay-restricted strategy $\astrat \in \drstrats{\semMDP{\PTA}}$
and a finite region path $\afinpathreg \in \mfinpaths{\regMDP{\PTA}}$,
there exists at most one finite path $\afinpath \in \mfinpaths{\astrat}$
such that $\afinpath \inregpath \afinpathreg$.
\end{lemma}
\begin{proof}
Let $\afinpathreg = (\aloc_0,\acec_0)(\anothercec_0,\apedge_0)
(\aloc_1,\acec_1)(\anothercec_1,\apedge_1)
\cdots
(\anothercec_{n-1},\apedge_{n-1})
(\aloc_n,\acec_n)$.
For $0 \leq i \leq n$, we write $\prefof{\afinpathreg}{i}$ as the $(i+1)$-th prefix of $\afinpathreg$,
i.e., $\prefof{\afinpathreg}{0} = (\aloc_0,\acec_0)$
and 
$\prefof{\afinpathreg}{i} = (\aloc_0,\acec_0)(\anothercec_0,\apedge_0)
\cdots
(\anothercec_{i-1},\apedge_{i-1})
(\aloc_i,\acec_i)$ for $0 < i \leq n$.
We proceed by attempting to construct $\afinpath \in \mfinpaths{\astrat}$
such that $\afinpath \inregpath \afinpathreg$ 
by attempting to construct progressively longer paths 
$\afinpath_0, \afinpath_1, \cdots, \afinpath_n$
in $\mfinpaths{\astrat}$ such that  
(1)~$\afinpath_i$ extends $\afinpath_{i-1}$ with one transition, for each $0 < i \leq n$,
and
(2)~$\afinpath_i \inregpath \prefof{\afinpathreg}{i}$ for each $0 \leq i \leq n$.
If the construction of paths $\afinpath_0, \afinpath_1, \cdots, \afinpath_n$
succeeds, then we let $\afinpath = \afinpath_n$.

Consider $i=0$. 
Then $\prefof{\afinpathreg}{0} = (\aloc_0,\acec_0)$.
Given that $\afinpathreg \in \mfinpaths{\regMDP{\PTA}}$,
we have that $(\aloc_0,\acec_0)$ is the initial state of $\regMDP{\PTA}$,
i.e., $(\aloc_0,\acec_0) = \reginit$.
Recalling that $(\locinit,\bzero)$ is the initial state of $\semMDP{\PTA}$,
we let $\afinpath_0 = (\locinit,\bzero)$.
Given that, trivially, $\bzero \in \cecof{\bzero}$,
we have $\afinpath_0 \inregpath \prefof{\afinpathreg}{0}$. 

Consider $0 < i \leq n$.
Assume that we have constructed $\afinpath_{i-1}$ 
such that $\afinpath_{i-1} \inregpath \prefof{\afinpathreg}{i-1}$;
we now attempt to construct $\afinpath_i$.
In order to have $\afinpath_i \inregpath \prefof{\afinpathreg}{i}$,
we need to identify $(\adelay_{i-1},\apedge_{i-1}) \in \support(\astrat(\afinpath_{i-1}))$
such that $\valof{\last(\afinpath_{i-1})}{+}\adelay_{i-1} \in \anothercec_{i-1}$ and $\aval_i \in \acec_i$.
Given that $\astrat$ is a delay-restricted strategy,
there exists at most one $(\adelay_{i-1},\apedge_{i-1}) \in \support(\astrat(\afinpath_{i-1}))$
such that $\valof{\last(\afinpath_{i-1})}{+}\adelay_{i-1} \in \anothercec_{i-1}$.

In the case that such 
$(\adelay_{i-1},\apedge_{i-1}) \in \support(\astrat(\afinpath_{i-1}))$ 
does not exist,
there is no possibility of extending $\afinpath_{i-1}$ to eventually obtain a path 
$\afinpath \in \mfinpaths{\astrat}$
such that $\afinpath \inregpath \afinpathreg$,
hence we are done.

In the case that such $(\adelay_{i-1},\apedge_{i-1}) \in \support(\astrat(\afinpath_{i-1}))$ exists,
from the fact that $\anothercec_{i-1}[\aclockset:=0] = \acec_i$ 
for some $\aclockset \subseteq \clocks$ such that $\distof{\apedge_{i-1}}(\aclockset,\aloc_i) > 0$,
and that $\valof{\last(\afinpath_{i-1})}{+}\adelay_{i-1} \in \anothercec_{i-1}$,
we have that $(\valof{\last(\afinpath_{i-1})}{+}\adelay_{i-1})[\aclockset:=0] \in \acec_i$;
hence we let $\aval_i = (\valof{\last(\afinpath_{i-1})}{+}\adelay_{i-1})[\aclockset:=0]$.
\qed
\end{proof}

\subsection{Proof of \lemref{lem:dr_finpaths}}

\begin{proof}
We first define the delay-restricted strategy $\adrstrat \in \drstrats{\semMDP{\PTA}}$
and then show that $\probfin{\astrat}(\dfinpathsfromregpath{\astrat}{\afinpathreg})
=
\probfin{\adrstrat}(\dfinpathsfromregpath{\adrstrat}{\afinpathreg})$
holds for any finite region path $\afinpathreg$ of $\regMDP{\PTA}$.

We introduce the following notation:
given a finite path $\afinpath \in \mfinpaths{\astrat}$,
we abbreviate $\valof{\last(\afinpath)}$ to $\vallast{\afinpath}$.

Let $\afinpath \in \mfinpaths{\semMDP{\PTA}}$,
and recall that $\afinpathregfrom{\afinpath}$ denotes the unique finite region path such that
$\afinpath \inregpath \afinpathregfrom{\afinpath}$.
We will define $\adrstrat(\afinpath)$ based on the choices of $\astrat$ from 
finite paths in $\dfinpathsfromregpath{\astrat}{\afinpathregfrom{\afinpath}}$.
Let $\last(\afinpathregfrom{\afinpath}) = (\aloc,\acec)$.
Consider $(\anothercec,\apedge) \in \regactionsof{\aloc,\acec}$.
Let $\choices{\afinpath'}{\astrat}{(\anothercec,\apedge)}$
be the actions chosen by $\astrat$ 
after finite path $\afinpath' \in \dfinpathsfromregpath{\astrat}{\afinpathregfrom{\afinpath}}$
that correspond to the selection of 
\begin{inparaenum}[(i)]
\item
a time delay resulting in a clock valuation in $\anothercec$, and
\item
the probabilistic edge $\apedge$.
\end{inparaenum}
Formally:
\[
\choices{\afinpath'}{\astrat}{(\anothercec,\apedge)}
= \set{ (\adelay,\apedge) \in \support(\astrat(\afinpath')) 
\setsep 
\vallast{\afinpath'}{+}\adelay \in \anothercec } \; .
\]
Note that $\set{ \choices{\afinpath'}{\astrat}{(\anothercec,\apedge)} 
\setsep
(\anothercec,\apedge) \in \regactionsof{\aloc,\acec}
}$
is a partition of $\support(\astrat(\afinpath'))$.
For each $\anothercec \in \cecs$ such that 
there exists $\apedge \in \pedges$ for which
(1)~$(\anothercec,\apedge) \in \regactionsof{\aloc,\acec}$
and (2)~$\choices{\afinpath'}{\astrat}{(\anothercec,\apedge)} \neq \emptyset$,
fix $\adrdelay{\anothercec} \in \nnr$ 
such that $\vallast{\afinpath'}{+}\adrdelay{\anothercec} \in \anothercec$.
Note that, from $(\anothercec,\apedge) \in \regactionsof{\aloc,\acec}$, 
we have $\anothercec \vinz \grdof{\apedge}$, 
and hence $\vallast{\afinpath}{+}\adrdelay{\anothercec} \vinz \grdof{\apedge}$.
Then we let:
\[
\adrstrat(\afinpath)(\adrdelay{\anothercec},\apedge)
= 
\frac{1}{\probfin{\astrat}(\dfinpathsfromregpath{\astrat}{\afinpathregfrom{\afinpath}})}
\sum_{\afinpath' \in \dfinpathsfromregpath{\astrat}{\afinpathregfrom{\afinpath}}}
\probfin{\astrat}(\afinpath')
\sum_{(\adelay,\apedge) \in \choices{\afinpath'}{\astrat}{(\anothercec,\apedge)}}
\astrat(\afinpath')(\adelay,\apedge) \; .
\]
We observe that $\adrstrat(\afinpath)$ is a distribution:
\begin{eqnarray*}
& & 
\sum_{(\adelay,\apedge) \in \support(\adrstrat(\afinpath))}
\adrstrat(\afinpath)(\adelay,\apedge)
\\
& \geq & 
\sum_{(\anothercec,\apedge) \in \regactionsof{\aloc,\acec}}
\adrstrat(\afinpath)(\adrdelay{\anothercec},\apedge)
\\
& = & 
\sum_{(\anothercec,\apedge) \in \regactionsof{\aloc,\acec}}
\frac{1}{\probfin{\astrat}(\dfinpathsfromregpath{\astrat}{\afinpathregfrom{\afinpath}})}
\sum_{\afinpath' \in \dfinpathsfromregpath{\astrat}{\afinpathregfrom{\afinpath}}}
\probfin{\astrat}(\afinpath')
\sum_{(\adelay,\apedge) \in \choices{\afinpath'}{\astrat}{(\anothercec,\apedge)}}
\astrat(\afinpath')(\adelay,\apedge)
\\
& = & 
\frac{1}{\probfin{\astrat}(\dfinpathsfromregpath{\astrat}{\afinpathregfrom{\afinpath}})}
\sum_{\afinpath' \in \dfinpathsfromregpath{\astrat}{\afinpathregfrom{\afinpath}}}
\probfin{\astrat}(\afinpath')
\sum_{(\anothercec,\apedge) \in \regactionsof{\aloc,\acec}}
\sum_{(\adelay,\apedge) \in \choices{\afinpath'}{\astrat}{(\anothercec,\apedge)}}
\astrat(\afinpath')(\adelay,\apedge)
\\
& = & 
\frac{1}{\probfin{\astrat}(\dfinpathsfromregpath{\astrat}{\afinpathregfrom{\afinpath}})}
\sum_{\afinpath' \in \dfinpathsfromregpath{\astrat}{\afinpathregfrom{\afinpath}}}
\probfin{\astrat}(\afinpath')
\\
& = & 
1 \; ,
\end{eqnarray*}\sloppypar{\noindent where the penultimate step follows from the fact that 
$\set{ \choices{\afinpath'}{\astrat}{(\anothercec,\apedge)} 
\setsep
(\anothercec,\apedge) \in \regactionsof{\aloc,\acec}
}$
is a partition of $\support(\astrat(\afinpath'))$,
and the final step follows from the definition of 
$\probfin{\astrat}(\dfinpathsfromregpath{\astrat}{\afinpathregfrom{\afinpath}})$.
We also note that the definition of $\adrstrat(\afinpath)$
is coherent with a delay-restricted strategy:
given $(\anothercec,\apedge) \in \regactionsof{\aloc,\acec}$,
there is at most one $(\adelay,\apedge) \in \support(\adrstrat(\afinpath))$
such that 
$\vallast{\afinpath}{+}\adelay \in \anothercec$.
}

To complete the proof,
we need to show that 
$\probfin{\astrat}(\dfinpathsfromregpath{\astrat}{\afinpathreg})
=
\probfin{\adrstrat}(\dfinpathsfromregpath{\adrstrat}{\afinpathreg})$
for an arbitrary finite region path $\afinpathreg$ of $\regMDP{\PTA}$,
We proceed by induction on the length of the considered finite region paths.
First consider the finite region path comprising the initial state
$\afinpathreg = \reginit$ of $\regMDP{\PTA}$.
Then, trivially, we have $\dfinpathsfromregpath{\astrat}{\afinpathreg}
= \dfinpathsfromregpath{\adrstrat}{\afinpathreg} = \set{ (\locinit, \bzero) }$,
for which 
$\probfin{\astrat}(\set{ (\locinit, \bzero) })
=
\probfin{\adrstrat}(\set{ (\locinit, \bzero) })
= 1$.

Let $i \geq 1$, and 
consider the finite region path $\afinpathreg \in \mfinpaths{\regMDP{\PTA}}$ 
of length $i-1$,
and the finite region path 
$\afinpathreg (\anothercec,\apedge) (\aloc,\acec) \in \mfinpaths{\regMDP{\PTA}}$,
which extends $\afinpathreg$.
%
By induction, we have established that
$\probfin{\astrat}(\dfinpathsfromregpath{\astrat}{\afinpathreg})
=
\probfin{\adrstrat}(\dfinpathsfromregpath{\adrstrat}{\afinpathreg})$.
If $\dfinpathsfromregpath{\astrat}{\afinpathreg} = \emptyset$,
i.e., $\afinpathreg$ does not correspond to any finite path generated by $\astrat$,
then, by construction of $\adrstrat$,
we have $\dfinpathsfromregpath{\adrstrat}{\afinpathreg} = \emptyset$.
Furthermore, no extension of $\afinpathreg$
can have a corresponding finite path generated by $\astrat$ or $\adrstrat$,
hence
$\dfinpathsfromregpath{\astrat}{\afinpathreg (\anothercec,\apedge) (\aloc,\acec)} = \emptyset$
and 
$\dfinpathsfromregpath{\adrstrat}{\afinpathreg (\anothercec,\apedge) (\aloc,\acec)} = \emptyset$,
from which we conclude that 
$\probfin{\astrat}(\dfinpathsfromregpath{\astrat}{\afinpathreg (\anothercec,\apedge) (\aloc,\acec)})
=
\probfin{\adrstrat}(\dfinpathsfromregpath{\adrstrat}{\afinpathreg (\anothercec,\apedge) (\aloc,\acec)})
= 0$.

Now consider the case in which $\dfinpathsfromregpath{\astrat}{\afinpathreg} \neq \emptyset$.
Let $\adrfinpath \in \mfinpaths{\adrstrat}$ be the unique finite path of $\adrstrat$
such that $\adrfinpath \inregpath \afinpathreg$
(i.e., $\dfinpathsfromregpath{\adrstrat}{\afinpathreg} = \set{ \adrfinpath }$).
There are two sub-cases.
If 
$\dfinpathsfromregpath{\astrat}{\afinpathreg (\anothercec,\apedge) (\aloc,\acec)} = \emptyset$,
then $\choices{\afinpath}{\astrat}{(\anothercec,\apedge)} = \emptyset$
for each $\afinpath \in \mfinpaths{\semMDP{\PTA}}$ such that 
$\afinpath \inregpath \afinpathreg$.
From the construction of $\adrstrat$,
there does not exist any $\adelay \in \nnr$ such that 
$(\adelay,\apedge) \in \support(\adrstrat(\adrfinpath))$
and $\vallast{\adrfinpath}{+}\adelay \in \anothercec$.
Hence 
$\dfinpathsfromregpath{\adrstrat}{\afinpathreg (\anothercec,\apedge) (\aloc,\acec)} = \emptyset$,
and therefore
$\probfin{\astrat}(\dfinpathsfromregpath{\astrat}{\afinpathreg (\anothercec,\apedge) (\aloc,\acec)})
=
\probfin{\adrstrat}(\dfinpathsfromregpath{\adrstrat}{\afinpathreg (\anothercec,\apedge) (\aloc,\acec)})
= 0$.

We now consider the second sub-case,
for which 
$\dfinpathsfromregpath{\astrat}{\afinpathreg (\anothercec,\apedge) (\aloc,\acec)} \neq \emptyset$.
First note that, 
from the fact that $\afinpathreg (\anothercec,\apedge) (\aloc,\acec)$
is a finite region path of $\regMDP{\PTA}$,
we have that  
$\regtrans(\last(\afinpathreg),(\anothercec,\apedge))(\aloc,\acec) > 0$,
which in turn means that
there exists $\aclockset \in \reset{\anothercec}{\acec}$
such that $\distof{\apedge}(\aclockset,\aloc) > 0$.
Next, from the construction of $\adrstrat$, we observe that
$(\adrdelay{\anothercec},\apedge) \in \support(\adrstrat(\adrfinpath))$.
Let $\adrval = \vallast{\adrfinpath}$.
We use $\adrvalof{\adrdelay{\anothercec}}{\acec}$ 
to denote the clock valuation in $\acec$ obtained 
by letting $\adrdelay{\anothercec}$ time elapse from $\adrval$
and then resetting clocks in an arbitrarily chosen set in $\reset{\anothercec}{\acec}$ to $0$.
Formally, 
$\adrvalof{\adrdelay{\anothercec}}{\acec} = (\adrval{+}\adrdelay{\anothercec})[\aclockset:=0]$
for some $\aclockset \in \reset{\anothercec}{\acec}$.
The fact that the choice of $\aclockset \in \reset{\anothercec}{\acec}$
can be made arbitrarily arises from 
the fact that $\adrvalof{\adrdelay{\anothercec}}{\acec} \in \acec$
and from \lemref{lem:reset_cor}, i.e., from 
$\reset{\adrval{+}\adrdelay{\anothercec}}{\adrvalof{\adrdelay{\anothercec}}{\acec}} 
= \reset{\anothercec}{\acec}$:
it holds that
$(\adrval{+}\adrdelay{\anothercec})[\aclockset:=0] 
= (\adrval{+}\adrdelay{\anothercec})[\aclockset':=0]
= \adrvalof{\adrdelay{\anothercec}}{\acec}$
for any $\aclockset,\aclockset' \in \reset{\anothercec}{\acec}$.
Now,
recalling that 
there exists $\aclockset \in \reset{\anothercec}{\acec}$
such that $\distof{\apedge}(\aclockset,\aloc) > 0$,
and from $\reset{\adrval{+}\adrdelay{\anothercec}}{\adrvalof{\adrdelay{\anothercec}}{\acec}} 
= \reset{\anothercec}{\acec}$ by \lemref{lem:reset_cor},
we have that there exists 
$\aclockset \in \reset{\adrval{+}\adrdelay{\anothercec}}{\adrvalof{\adrdelay{\anothercec}}{\acec}}$
such that $\distof{\apedge}(\aclockset,\aloc) > 0$.
This fact then means that 
$\mdptrans(\last(\adrfinpath),(\adrdelay{\anothercec},\apedge))
(\aloc,\adrvalof{\adrdelay{\anothercec}}{\acec}) > 0$.
Recall that the definition of $\adrstrat$ specifies that
$\adrdelay{\anothercec}$ is the \emph{unique} time delay 
such that $\adrstrat(\adrdelay{\anothercec},\apedge)>0$ 
and $\adrval{+}\adrdelay{\anothercec} \in \anothercec$.
Furthermore, recall that $\adrvalof{\adrdelay{\anothercec}}{\acec}$
is the \emph{unique} clock valuation in $\acec$ obtained by 
letting $\adrdelay{\anothercec}$ time elapse from $\adrval$
and then resetting clocks in some set in $\reset{\anothercec}{\acec}$ to $0$.
Combining these two facts 
with the fact that 
$\mdptrans(\last(\adrfinpath),(\adrdelay{\anothercec},\apedge))
(\aloc,\adrvalof{\adrdelay{\anothercec}}{\acec}) > 0$,
we conclude that 
$\dfinpathsfromregpath{\adrstrat}{\afinpathreg (\anothercec,\apedge) (\aloc,\acec)}
= \set{ \adrfinpath (\adrdelay{\anothercec},\apedge) (\aloc, \adrvalof{\adrdelay{\anothercec}}{\acec}) }$,
and thus 
$\probfin{\adrstrat}(\dfinpathsfromregpath{\adrstrat}{\afinpathreg (\anothercec,\apedge) (\aloc,\acec)})
= \probfin{\adrstrat}(\adrfinpath (\adrdelay{\anothercec},\apedge) (\aloc, \adrvalof{\adrdelay{\anothercec}}{\acec}))$.
Our task therefore becomes that of showing
$\probfin{\adrstrat}(\adrfinpath (\adrdelay{\anothercec},\apedge) (\aloc, \adrvalof{\adrdelay{\anothercec}}{\acec}))
=
\probfin{\astrat}(\dfinpathsfromregpath{\astrat}{\afinpathreg (\anothercec,\apedge) (\aloc,\acec)})$.

From the definition of $\probfin{\adrstrat}$,
and recalling that $\reset{\adrval{+}\adrdelay{\anothercec}}{\adrvalof{\adrdelay{\anothercec}}{\acec}} 
= \reset{\anothercec}{\acec}$ by \lemref{lem:reset_cor}, 
we have:
\begin{eqnarray*}
\probfin{\adrstrat}(\adrfinpath (\adrdelay{\anothercec},\apedge) (\aloc, \adrvalof{\adrdelay{\anothercec}}{\acec}))
& = &
\probfin{\adrstrat}(\adrfinpath)
\cdot 
\adrstrat(\adrfinpath)(\adrdelay{\anothercec},\apedge) 
\cdot
\sum_{\aclockset \in
\reset{\vallast{\adrfinpath}{+}\adrdelay{\anothercec}}{\adrvalof{\adrdelay{\anothercec}}{\acec}}}
\distof{\apedge}(\aclockset,\aloc)
\\
& = &
\probfin{\adrstrat}(\adrfinpath)
\cdot 
\adrstrat(\adrfinpath)(\adrdelay{\anothercec},\apedge) 
\cdot
\sum_{\aclockset \in
\reset{\anothercec}{\acec}}
\distof{\apedge}(\aclockset,\aloc)
\\
& = &
\probfin{\astrat}(\dfinpathsfromregpath{\astrat}{\afinpathreg})
\cdot 
\adrstrat(\adrfinpath)(\adrdelay{\anothercec},\apedge) 
\cdot
\sum_{\aclockset \in
\reset{\anothercec}{\acec}}
\distof{\apedge}(\aclockset,\aloc) \; ,
\end{eqnarray*}
where the final step follows by induction.

Now consider 
$\probfin{\astrat}(\dfinpathsfromregpath{\astrat}{\afinpathreg})
\cdot 
\adrstrat(\adrfinpath)(\adrdelay{\anothercec},\apedge)$.
From the definition of $\adrstrat$
(and recalling that $\adrfinpath \inregpath \afinpathreg$),
we have:
\begin{eqnarray*}
& &
\probfin{\astrat}(\dfinpathsfromregpath{\astrat}{\afinpathreg})
\cdot 
\adrstrat(\adrfinpath)(\adrdelay{\anothercec},\apedge)
\\
& = &
\probfin{\astrat}(\dfinpathsfromregpath{\astrat}{\afinpathreg})
\cdot 
\frac{1}{\probfin{\astrat}(\dfinpathsfromregpath{\astrat}{\afinpathreg})}
\sum_{\afinpath \in \dfinpathsfromregpath{\astrat}{\afinpathreg}}
\probfin{\astrat}(\afinpath)
\sum_{(\adelay,\apedge) \in \choices{\afinpath}{\astrat}{(\anothercec,\apedge)}}
\astrat(\afinpath)(\adelay,\apedge)
\\
& = &
\sum_{\afinpath \in \dfinpathsfromregpath{\astrat}{\afinpathreg}}
\probfin{\astrat}(\afinpath)
\sum_{(\adelay,\apedge) \in \choices{\afinpath}{\astrat}{(\anothercec,\apedge)}}
\astrat(\afinpath)(\adelay,\apedge) \; .
\end{eqnarray*}
Substituting this in the previous equation allows us to obtain:
\begin{eqnarray*}
& & 
\probfin{\adrstrat}(\adrfinpath (\adrdelay{\anothercec},\apedge) (\aloc, \adrvalof{\adrdelay{\anothercec}}{\acec}))
\\
& = &
\sum_{\afinpath \in \dfinpathsfromregpath{\astrat}{\afinpathreg}}
\probfin{\astrat}(\afinpath)
\sum_{(\adelay,\apedge) \in \choices{\afinpath}{\astrat}{(\anothercec,\apedge)}}
\astrat(\afinpath)(\adelay,\apedge)
\sum_{\aclockset \in
\reset{\anothercec}{\acec}}
\distof{\apedge}(\aclockset,\aloc) \; .
\end{eqnarray*}
For $\afinpath \in \dfinpathsfromregpath{\astrat}{\afinpathreg}$
and $\adelay \in \nnr$ such that $\vallast{\afinpath}{+}\adelay \in \anothercec$,
let $\avalof{\afinpath,\adelay}{\acec} = (\vallast{\afinpath}{+}\adelay)[\aclockset:=0]$
be the clock valuation obtained from the clock valuation 
of the last state of the finite path $\afinpath$
after letting time elapse and resetting clocks in $\aclockset$ to $0$,
for some arbitrarily chosen $\aclockset \in \reset{\anothercec}{\acec}$.
Note that $\avalof{\afinpath,\adelay}{\acec} \in \acec$.
From \lemref{lem:reset_cor}, we have 
$\reset{\anothercec}{\acec} 
= \reset{\vallast{\afinpath}{+}\adelay}{\avalof{\afinpath,\adelay}{\acec}}$.
Hence:
\begin{eqnarray*}
& & 
\probfin{\adrstrat}(\adrfinpath (\adrdelay{\anothercec},\apedge) (\aloc, \adrvalof{\adrdelay{\anothercec}}{\acec}))
\\
& = &
\sum_{\afinpath \in \dfinpathsfromregpath{\astrat}{\afinpathreg}}
\probfin{\astrat}(\afinpath)
\sum_{(\adelay,\apedge) \in \choices{\afinpath}{\astrat}{(\anothercec,\apedge)}}
\astrat(\afinpath)(\adelay,\apedge)
\sum_{\aclockset \in
\reset{\vallast{\afinpath}{+}\adelay}{\avalof{\afinpath,\adelay}{\acec}}}
\distof{\apedge}(\aclockset,\aloc) \; .
\end{eqnarray*}
From the definition of $\choices{\afinpath}{\astrat}{(\anothercec,\apedge)}$,
we obtain:
\begin{eqnarray*}
& & 
\probfin{\adrstrat}(\adrfinpath (\adrdelay{\anothercec},\apedge) (\aloc, \adrvalof{\adrdelay{\anothercec}}{\acec}))
\\
& = &
\sum_{\afinpath \in \dfinpathsfromregpath{\astrat}{\afinpathreg}}
\probfin{\astrat}(\afinpath)
\sum_{\substack{(\adelay,\apedge) \in \support(\astrat(\afinpath))\\ \mbox{ s.t. } \vallast{\afinpath}{+}\adelay \in \anothercec}}
\astrat(\afinpath)(\adelay,\apedge)
\sum_{\aclockset \in
\reset{\vallast{\afinpath}{+}\adelay}{\avalof{\afinpath,\adelay}{\acec}}}
\distof{\apedge}(\aclockset,\aloc)
\\
& = &
\sum_{\afinpath \in \dfinpathsfromregpath{\astrat}{\afinpathreg}}
\sum_{\substack{(\adelay,\apedge) \in \support(\astrat(\afinpath))\\ \mbox{ s.t. } \vallast{\afinpath}{+}\adelay \in \anothercec}}
\probfin{\astrat}(\afinpath)
\cdot 
\astrat(\afinpath)(\adelay,\apedge)
\sum_{\aclockset \in
\reset{\vallast{\afinpath}{+}\adelay}{\avalof{\afinpath,\adelay}{\acec}}}
\distof{\apedge}(\aclockset,\aloc)
\\
& = &
\sum_{\afinpath (\adelay,\apedge) (\aloc,\avalof{\adelay,\apedge}{\aloc,\acec}) 
\in 
\dfinpathsfromregpath{\astrat}{\afinpathreg (\anothercec,\apedge) (\aloc,\acec)}}
\probfin{\astrat}(\afinpath (\adelay,\apedge) (\aloc,\avalof{\afinpath,\adelay}{\acec}))
\\
& = &
\probfin{\astrat}(\dfinpathsfromregpath{\astrat}{\afinpathreg (\anothercec,\apedge) (\aloc,\acec)})
\; .
\end{eqnarray*}

Hence we have shown that 
$\probfin{\adrstrat}(\dfinpathsfromregpath{\adrstrat}{\afinpathreg (\anothercec,\apedge) (\aloc,\acec)})
=
\probfin{\astrat}(\dfinpathsfromregpath{\astrat}{\afinpathreg (\anothercec,\apedge) (\aloc,\acec)})$,
completing the proof.
\qed
\end{proof}

\subsection{Proof of \lemref{lem:dr_mof}}

\begin{proof}
Given $\astrat \in \strats{\semMDP{\PTA}}$ such that 
$\objsat{(\semMDP{\PTA},\mdplabof{\semMDP{\PTA}})}{\astrat}{\mof}$,
we construct $\adrstrat \in \drstrats{\semMDP{\PTA}}$ 
following the approach of the proof of \lemref{lem:dr_finpaths}.
It remains to show that $\objsat{(\semMDP{\PTA},\mdplabof{\semMDP{\PTA}})}{\adrstrat}{\mof}$.
Let $\mof = \bigwedge_{i=1}^n \anobj_i(\anevent_i)$.

First we consider the case of a subformula $\anobj_i(\anevent_i)$ 
for which $\anobj_i$ is $\sure$.
Note that $\adrstrat$ differs from $\astrat$ in the following way:
randomized choice between
different delays obtaining clock valuations within the same clock equivalence class 
(and then taking the same probabilistic edge) in $\astrat$
is represented in $\adrstrat$ by the choice of a \emph{single} delay
obtaining a clock valuation 
in the aforementioned clock equivalence class.
Hence the sequences of locations visited (i.e., set of traces exhibited) 
by $\astrat$ and $\adrstrat$ are identical.
Formally, we have 
$\set{ \trace{\mdplabof{\semMDP{\PTA}}}{\apath} \setsep \apath \in \mpaths{\astrat} }
= \set{ \trace{\mdplabof{\semMDP{\PTA}}}{\adrpath} \setsep \adrpath \in \mpaths{\adrstrat} }$,
from which we conclude that 
$\objsat{(\semMDP{\PTA},\mdplabof{\semMDP{\PTA}})}{\astrat}{\sure(\anevent_i)}$
implies that $\objsat{(\semMDP{\PTA},\mdplabof{\semMDP{\PTA}})}{\adrstrat}{\sure(\anevent_i)}$.

Now consider the case of $\anobj_i(\anevent_i)$ 
for which $\anobj_i$ is of the form $\probthresh{\thresh}{\lambda}$.
Our reasoning is similar to that used in the proof of Lemma~10.66 in \cite{BK08}.
Consider a finite sequence $\alabseq$ of locations.
Note that $\alabseq$ corresponds to a finite set of finite region paths of $\regMDP{\PTA}$,
i.e., the set of finite region paths visiting locations in the order specified by $\alabseq$.
From \lemref{lem:dr_finpaths},
we have that the sum of probabilities assigned by $\astrat$ and $\adrstrat$ to 
finite paths exhibiting $\alabseq$ are equal.
From this fact, 
and standard reasoning about the extension of functions assigning the probability of finite paths
to probability measures over sets of infinite paths,
we know that the probabilities assigned by $\dprob{\astrat}$ and $\dprob{\adrstrat}$
to the sets of infinite paths with prefixes following the sequence $\alabseq$ are equal.
This means that, for any event $\anevent$
(a measurable set of infinite sequences of locations),
we have that $\dprob{\astrat}(\dpathsfromstrat{\astrat}{\anevent})
= \dprob{\adrstrat}(\dpathsfromstrat{\adrstrat}{\anevent})$.
Hence if $\objsat{(\semMDP{\PTA},\mdplabof{\semMDP{\PTA}})}{\astrat}{\probthresh{\thresh}{\lambda}(\anevent_i)}$
then $\objsat{(\semMDP{\PTA},\mdplabof{\semMDP{\PTA}})}{\adrstrat}{\probthresh{\thresh}{\lambda}(\anevent_i)}$.\footnote{
Note that, in this case, and the case in which $\anobj_i$ is $\sure$,
the reverse implication is also true, 
but we do not need that direction for our results.
}
\qed
\end{proof}

\subsection{Proof of \lemref{lem:dr_to_reg}}

\begin{proof}
Let $\adrstrat \in \drstrats{\semMDP{\PTA}}$ 
be a delay-restricted strategy.
We define the strategy $\aregstratof{\adrstrat}$ of $\regMDP{\PTA}$.
Let $\afinpath \in \mfinpaths{\adrstrat}$,
and recall that $\afinpathregfrom{\afinpath}$
denotes the unique finite region path of $\regMDP{\PTA}$
that corresponds to $\afinpath$.
Then we define $\aregstratof{\adrstrat}(\afinpathregfrom{\afinpath})$ in the following way.
Given that $\adrstrat$ is a delay-restricted strategy,
for $(\adelay,\apedge),(\adelay',\apedge') \in \support(\adrstrat(\afinpath))$
such that $\apedge = \apedge'$,
we have $\cecof{\vallast{\afinpath}+\adelay} \neq \cecof{\vallast{\afinpath}+\adelay'}$.
This means that we can define the injective function 
$\aninj: \support(\adrstrat(\afinpath)) \ra \regactions(\last(\afinpathregfrom{\afinpath}))$
defined by $\aninj(\adelay,\apedge) = (\cecof{\vallast{\afinpath}+\adelay}, \apedge)$.
Then, for each $(\adelay,\apedge) \in \support(\adrstrat(\afinpath))$,  
we let $\aregstratof{\adrstrat}(\afinpathregfrom{\afinpath})(\aninj(\adelay,\apedge))
= \adrstrat(\afinpath)(\adelay,\apedge)$.
For each $(\anothercec,\apedge) \in \regactions(\last(\afinpathregfrom{\afinpath}))$
such that there does not exist $(\adelay,\apedge) \in \support(\adrstrat(\afinpath))$
with $(\anothercec,\apedge) = \aninj(\adelay,\apedge)$,
we let $\aregstratof{\adrstrat}(\afinpathregfrom{\afinpath})(\anothercec,\apedge) = 0$.
Observe that $\aregstratof{\adrstrat}(\afinpathregfrom{\afinpath})$ is a distribution.
For all finite region paths $\afinpathreg \in \mfinpaths{\regMDP{\PTA}}$
such that there does not exist any finite path $\afinpath \in \mfinpaths{\adrstrat}$
such that $\afinpathregfrom{\afinpath} = \afinpathreg$,
the choice made in $\aregstratof{\adrstrat}(\afinpathreg)$ is arbitrary.
This completes the definition of $\aregstratof{\adrstrat}$.

Recall that the labelling function of 
the labelled semantic MDP $(\semMDP{\PTA},\mdplabof{\semMDP{\PTA}})$
(respectively, of the labelled region-graph MDP $(\regMDP{\PTA},\mdplabof{\regMDP{\PTA}})$) 
assigns to a state (respectively, a region) its location.
Furthermore, 
recall that the labelling function of the labelled MC corresponding to a strategy
assigns to an MC state, which is a finite path of the strategy,
the location of the final MDP state of the path.
We now show that the labelled MCs 
$(\adtmcparaone{\adrstrat},\dtmclabfrom{\mdplabof{\semMDP{\PTA}}})$ 
and 
$(\adtmcparaone{\aregstratof{\adrstrat}},\dtmclabfrom{\mdplabof{\regMDP{\PTA}}})$
are isomorphic.
We propose the bijection $\abij: \mfinpaths{\adrstrat} \ra \mfinpaths{\aregstratof{\adrstrat}}$
defined by $\abij(\afinpath) = \afinpathregfrom{\afinpath}$
for all $\afinpath \in \mfinpaths{\adrstrat}$.
The fact that $\abij$ is a bijection follows from the definition of $\aregstratof{\adrstrat}$:
we have that $\abij$ maps to finite region paths of $\aregstratof{\adrstrat}$ only 
(i.e., given $\afinpath \in \mfinpaths{\adrstrat}$,
we have $\afinpathregfrom{\afinpath} \in \mfinpaths{\aregstratof{\adrstrat}}$),
and that all finite region paths of $\aregstratof{\adrstrat}$ 
have a corresponding finite path of $\adrstrat$
(i.e., for all $\afinpathreg \in \mfinpaths{\aregstratof{\adrstrat}}$
there exists $\afinpath \in \mfinpaths{\adrstrat}$ 
such that $\abij(\afinpath) = \afinpathreg$).
For part~(1) of the definition of isomorphic labelled MCs,
recall that the initial states of $\adtmcparaone{\adrstrat}$ 
and $\adtmcparaone{\aregstratof{\adrstrat}}$
are $(\locinit,\bzero)$ and $\reginit$, respectively.
Given that $(\locinit,\bzero)$ and $\reginit$, 
interpreted as paths with a single state and no transitions,
are such that $(\locinit,\bzero) \inregpath \reginit$,
i.e., $\afinpathregfrom{(\locinit,\bzero)} = \reginit$,
we obtain $\abij(\locinit,\bzero) = \reginit$.
Now we consider parts~(2a) and~(2b) of the definition of isomorphism. 
Consider $\afinpath \in \mfinpaths{\adrstrat}$ and 
$\afinpathreg \in \mfinpaths{\aregstratof{\adrstrat}}$
such that $\abij(\afinpath) = \afinpathreg$,
and hence $\afinpath \inregpath \afinpathreg$.
For (2a),
from $\afinpath \inregpath \afinpathreg$
we know that the last state along $\afinpath$ and the last region along $\afinpathreg$
have the same location,
from which we conclude that
$\dtmclabfrom{\mdplabof{\semMDP{\PTA}}}(\afinpath)
=
\dtmclabfrom{\mdplabof{\regMDP{\PTA}}}(\afinpathreg)$.
For (2b),
consider the finite path $\afinpath (\adelay,\apedge) (\aloc,\aval) \in \mfinpaths{\adrstrat}$.
To establish (2b), we need to show that:
\begin{eqnarray*}
\dtransof{\adrstrat}(\afinpath, \afinpath (\adelay,\apedge) (\aloc,\aval)) 
& = &
\dtransof{\aregstratof{\adrstrat}}(\afinpathreg, \abij(\afinpath (\adelay,\apedge) (\aloc,\aval))) \; .
\end{eqnarray*}
From the definitions of $\dtransof{\adrstrat}$ and $\semMDP{\PTA}$, we have:
\begin{eqnarray*}
\dtransof{\adrstrat}(\afinpath, \afinpath (\adelay,\apedge) (\aloc,\aval)) 
& = & 
\adrstrat(\afinpath)(\adelay,\apedge) \cdot \mtrans(\last(\afinpath),(\adelay,\apedge))(\aloc,\aval)
\\
& = &
\adrstrat(\afinpath)(\adelay,\apedge)
\sum_{\aclockset \in \reset{\vallast{\afinpath}{+}\adelay}{\aval}} \pd(\aclockset,\aloc) \; .
\end{eqnarray*}
Now note that,
from the definition of $\afinpathregfrom{\cdot}$ on finite paths,
the finite region path
$\afinpathregfrom{\afinpath (\adelay,\apedge) (\aloc,\aval)}$
corresponds to
$\afinpathregfrom{\afinpath} (\cecof{\vallast{\afinpath}{+}\adelay},\apedge) (\aloc,\cecof{\aval})$.
Hence we have $\abij(\afinpath (\adelay,\apedge) (\aloc,\aval)) = 
\afinpathregfrom{\afinpath} (\cecof{\vallast{\afinpath}{+}\adelay},\apedge) (\aloc,\cecof{\aval})$.
From the definitions of $\dtransof{\aregstratof{\adrstrat}}$ and $\regMDP{\PTA}$:
\begin{eqnarray*}
& &
\dtransof{\aregstratof{\adrstrat}}(\afinpathregfrom{\afinpath}, \afinpathregfrom{\afinpath} (\cecof{\vallast{\afinpath}{+}\adelay},\apedge)) (\aloc,\cecof{\aval})) 
\\
& = & 
\aregstratof{\adrstrat}(\afinpathregfrom{\afinpath})(\cecof{\vallast{\afinpath}{+}\adelay},\apedge) \cdot 
\regtrans(\last(\afinpathregfrom{\afinpath}),(\cecof{\vallast{\afinpath}{+}\adelay},\apedge)(\aloc,\cecof{\aval})
\\
& = &
\aregstratof{\adrstrat}(\afinpathregfrom{\afinpath})(\cecof{\vallast{\afinpath}{+}\adelay},\apedge)
\sum_{\aclockset \in \reset{\cecof{\vallast{\afinpath}{+}\adelay}}{\cecof{\aval}}} \pd(\aclockset,\aloc) \; .
\end{eqnarray*}
Hence, our task reduces to showing that:
\begin{eqnarray*}
\adrstrat(\afinpath)(\adelay,\apedge)
\sum_{\aclockset \in \reset{\vallast{\afinpath}{+}\adelay}{\aval}} \pd(\aclockset,\aloc)& = &
\aregstratof{\adrstrat}(\afinpathregfrom{\afinpath})(\cecof{\vallast{\afinpath}{+}\adelay},\apedge)
\sum_{\aclockset \in \reset{\cecof{\vallast{\afinpath}{+}\adelay}}{\cecof{\aval}}} \pd(\aclockset,\aloc) \; . \end{eqnarray*}
From the definition of $\aregstratof{\adrstrat}$,
we have $\adrstrat(\afinpath)(\adelay,\apedge) = \aregstratof{\adrstrat}(\afinpathregfrom{\afinpath})(\cecof{\vallast{\afinpath}{+}\adelay},\apedge)$.
Furthermore, from \lemref{lem:reset_cor}, 
we have $\reset{\vallast{\afinpath}{+}\adelay}{\aval} = \reset{\cecof{\vallast{\afinpath}{+}\adelay}}{\cecof{\aval}}$
and thus $\sum_{\aclockset \in \reset{\vallast{\afinpath}{+}\adelay}{\aval}} \pd(\aclockset,\aloc')
= 
\sum_{\aclockset \in \reset{\cecof{\vallast{\afinpath}{+}\adelay}}{\cecof{\aval}}} \pd(\aclockset,\aloc)$.
Therefore we conclude that 
$\dtransof{\adrstrat}(\afinpath, \afinpath (\adelay,\apedge) (\aloc,\aval)) 
= 
\dtransof{\aregstratof{\adrstrat}}(\afinpathregfrom{\afinpath}, \afinpathregfrom{\afinpath} (\cecof{\vallast{\afinpath}{+}\adelay},\apedge) (\aloc,\cecof{\aval}))$,
i.e., 
$\dtransof{\adrstrat}(\afinpath, \afinpath (\adelay,\apedge) (\aloc,\aval)) 
= 
\dtransof{\aregstratof{\adrstrat}}(\afinpathreg, \abij(\afinpath (\adelay,\apedge) (\aloc,\aval))$,
which satisfies condition~(2b) in the defnition of isomorphism.
Hence the labelled MCs 
$(\adtmcparaone{\adrstrat},\dtmclabfrom{\mdplabof{\semMDP{\PTA}}})$ 
and 
$(\adtmcparaone{\aregstratof{\adrstrat}},\dtmclabfrom{\mdplabof{\regMDP{\PTA}}})$
are isomorphic,
and we have completed the proof.
\qed
\end{proof}

\subsection{Proof of \lemref{lem:rg_to_sem}}

\begin{proof}
Let $\astrat \in \strats{\regMDP{\PTA}}$.
We define $\astrat' \in \strats{\semMDP{\PTA}}$ as follows.
Let $\afinpath \in \mfinpaths{\semMDP{\PTA}}$,
and consider the finite region path $\afinpathregfrom{\afinpath}$ of $\regMDP{\PTA}$.
For each $\anothercec \in \cecs$ such that 
there exists  
$(\anothercec,\apedge) \in \support(\astrat(\afinpathregfrom{\afinpath}))$,
we fix $\adelayparatwo{\afinpath}{\anothercec} \in \nnr$
such that $\vallast{\afinpath}{+}\adelayparatwo{\afinpath}{\anothercec} \in \anothercec$.
Then, for each $(\anothercec,\apedge) \in \support(\astrat(\afinpathregfrom{\afinpath}))$,
we let $\astrat'(\afinpath)(\adelayparatwo{\afinpath}{\anothercec},\apedge) 
= \astrat(\afinpathregfrom{\afinpath})(\anothercec,\apedge)$.
The fact that $(\adelayparatwo{\afinpath}{\anothercec},\apedge) \in \mactionsof{\last(\afinpath)}$,
which is necessary for $\astrat'$ to be well defined,
follows from two facts.
First, from 
$(\anothercec,\apedge) \in \support(\astrat(\afinpathregfrom{\afinpath})$,
we know that $\anothercec$ is a time successor of $\last(\afinpathregfrom{\afinpath})$,
hence the invariant condition of the location of $\last(\afinpath)$
 is satisfied as $\adelayparatwo{\afinpath}{\anothercec}$ time passes.
Second, $(\anothercec,\apedge) \in \support(\astrat(\afinpathregfrom{\afinpath})$
also implies that $\anothercec \vinz \grdof{\apedge}$,
hence $\vallast{\afinpath}{+}\adelayparatwo{\afinpath}{\anothercec} \in \grdof{\apedge}$.
To complete the definition of $\astrat'$,
we let $\astrat'(\afinpath)(\adelay,\apedge) = 0$
for all $(\adelay,\apedge) \in \mactionsof{\last(\afinpath)}$ not considered above.
This completes the definition of $\astrat'(\afinpath)$. 
Note that there is a 1-to-1 relationship between elements of 
$\support(\astrat(\afinpathregfrom{\afinpath}))$ and $\support(\astrat'(\afinpath))$,
and that for related elements 
$(\anothercec,\apedge) \in \support(\astrat(\afinpathregfrom{\afinpath}))$
and $(\adelayparatwo{\afinpath}{\anothercec},\apedge) \in \support(\astrat'(\afinpath))$,
we have $\astrat'(\afinpath)(\adelayparatwo{\afinpath}{\anothercec},\apedge) 
= \astrat(\afinpathregfrom{\afinpath})(\anothercec,\apedge)$.
As a consequence, we have that $\astrat'(\afinpath)$ is a distribution:
\begin{eqnarray*}
\sum_{(\adelay,\apedge) \in \support(\astrat'(\afinpath))}
\astrat'(\afinpath)(\adelay,\apedge)
& = & 
\sum_{(\anothercec,\apedge) \in \support(\astrat(\afinpathregfrom{\afinpath}))}
\astrat(\afinpathregfrom{\afinpath})(\anothercec,\apedge)
= 1 \; .
\end{eqnarray*}
Note also that $\astrat'$ is a delay-restricted strategy:
given $\afinpath \in \mfinpaths{\semMDP{\PTA}}$ 
and $(\anothercec,\apedge) \in \support(\astrat(\afinpathregfrom{\afinpath}))$,
there exists exactly one 
$(\adelayparatwo{\afinpath}{\anothercec},\apedge) \in \support(\astrat(\afinpathregfrom{\afinpath})$
such that $\adelayparatwo{\afinpath}{\anothercec} \in \anothercec$.
This fact will be useful in the remainder of the proof.

We now show that 
$(\adtmcparaone{\astrat},\dtmclabfrom{\mdplabof{\regMDP{\PTA}}})$
and
$(\adtmcparaone{\astrat'},\dtmclabfrom{\mdplabof{\semMDP{\PTA}}})$  
are isomorphic.
Recall that this requires a bijection on $\mfinpaths{\astrat}$
and $\mfinpaths{\astrat'}$, 
i.e., the sets of finite paths of $\adtmcparaone{\astrat}$
and $\adtmcparaone{\astrat'}$, respectively,
which satisfies certain properties.
Recall that, given that $\astrat'$ is a delay-restricted strategy,
\lemref{lem:drstrat_one_path} specifies that,
for each finite region path of $\regMDP{\PTA}$,
there exists at most one corresponding path of $\astrat'$.
Vice versa, for each finite path $\afinpath \in \mfinpaths{\astrat'}$ of $\astrat'$,
we have that $\afinpathregfrom{\afinpath}$ is a finite region path of $\astrat$,
i.e., $\afinpathregfrom{\afinpath} \in \mfinpaths{\astrat}$ from the following reasoning.
Writing $\afinpath = (\aloc_0,\aval_0)(\adelay_0,\apedge_0)
(\aloc_1,\aval_1)(\adelay_1,\apedge_1)
\cdots
(\adelay_{n-1},\apedge_{n-1})
(\aloc_n,\aval_n)$,
and writing $\prefof{\afinpath}{i}$ as the $(i+1)$-th prefix of $\afinpath$,
for $0 \leq i < n$
assume that we have established that 
$\afinpathregfrom{\prefof{\afinpath}{i}} \in \mfinpaths{\astrat}$
(for $i=0$, we have that $\afinpathregfrom{\prefof{\afinpath}{i}} = \reginit$,
which is trivially a finite region path in $\mfinpaths{\astrat}$).
Consider $(\adelay_i,\apedge_i)$:
from the fact that $\afinpath \in \mfinpaths{\astrat'}$,
we have $(\adelay_i,\apedge_i) \in \support(\astrat'(\prefof{\afinpath}{i}))$.
Then, from the 1-to-1 relationship between elements of 
$\support(\astrat(\afinpathregfrom{\prefof{\afinpath}{i}}))$ 
and $\support(\astrat'(\prefof{\afinpath}{i}))$ mentioned in the previous paragraph,
we have that 
$\afinpathregfrom{\prefof{\afinpath}{i}}(\anothercec,\apedge_i)(\aloc_{i+1},\acec_{i+1})
\in \mfinpaths{\astrat}$,
where $(\anothercec_i,\apedge_i)$ is the unique element of 
$\support(\astrat(\afinpathregfrom{\prefof{\afinpath}{i}}))$
related to $(\adelay_i,\apedge_i)$,
and $(\aloc_{i+1},\acec_{i+1})$ is the unique element of 
$\support(\regtrans(\last(\afinpathregfrom{\prefof{\afinpath}{i}}),(\anothercec_i,\apedge_i))))$
such that $\aval_{i+1} \in \acec_{i+1}$,
which exists from the same reasoning as given in the proof of \lemref{lem:drstrat_one_path}.

The above means that we can define the bijection 
$\abij: \mfinpaths{\astrat'} \ra \mfinpaths{\astrat}$
such that $\abij(\afinpath) = \afinpathregfrom{\afinpath}$.
The remainder of the proof then continues as does 
the analogous part of the proof of \lemref{lem:dr_to_reg}
(subsituting $\astrat$ for $\aregstratof{\adrstrat}$, and $\astrat'$ for $\adrstrat$)
and we omit the details.
\qed
\end{proof}

\end{document}